\documentclass[a4paper]{article}
\usepackage{amssymb}
\usepackage{amsmath,bm}
\usepackage{amsthm}
\usepackage{url}
\usepackage{cite}
\usepackage{subcaption}
\usepackage{float}
\usepackage{enumitem}
\usepackage[font=footnotesize]{caption}
\numberwithin{equation}{section}
\usepackage{graphicx}
\usepackage{xcolor}
\newcommand{\colorblue}[1]{#1}
\newtheorem{theorem}{Theorem}[section]
\newtheorem{prop}[theorem]{Proposition}
\newtheorem{lem}[theorem]{Lemma}

\newtheorem{remark}[theorem]{Remark}

\newcommand{\D}{\mathrm{d}}

\newcommand{\atanh}{{\, \mathrm{atanh}}}
\newcommand{\acoth}{{\, \mathrm{acoth}}}
\newcommand{\pp}[2]{ \frac{\partial #1}{\partial #2} }
\begin{document}
\title{Talanov dynamics and chirp-controlled rogue-wave formation for the vector nonlinear Schr\"odinger equation}
\author{G. Ortenzi$^{1,2}$\footnote{Corresponding author. E-mail: giovanni.ortenzi@unito.it}, G. Roberti$^{3}$, and M. Sommacal$^{3}$\\
$^{1}$ {\footnotesize Dipartimento di Matematica ``Giuseppe Peano'', Universit\`{a} di Torino, 10123 Torino, Italy}\\
$^{2}$ {\footnotesize INFN, Sezione di Torino, 10125 Torino, Italy}\\
$^{3}$ {\footnotesize School of Engineering, Physics and Mathematics, Northumbria University, NE1 8ST, United Kingdom}
}
\maketitle
\abstract{
We study the focusing vector nonlinear Schrödinger equation, or Manakov system, for a pair of coupled initial conditions with compactly supported, concave parabolic density profiles and phases depending quadratically on the spatial coordinate. We focus on the natural class of self-similar solutions generated by such profiles, whose dispersionless evolution we refer to as Talanov dynamics.  We extend the criterion for blow-up in finite time, first established for a single field, to the Manakov case. For arbitrary initial chirps in the two fields, we provide closed-form expressions for their evolution, and an explicit expression for the time of the gradient catastrophe, if it occurs, and we classify the effect of the coupling on the catastrophe. Depending on the initial chirps, the two fields may catastrophise, or relax, as each would if decoupled. However, coupling can also force the catastrophe in fields that would otherwise relax if decoupled. For identical initial supports, and in the presence of dispersion, we numerically show that the same dispersionless criterion determines, in the semiclassical regime, whether the fully dispersive solution relaxes or develops a rogue wave in each field, whose onset time is accurately predicted by the corresponding dispersionless catastrophe time, in direct analogy with the scalar case. As chirp and amplitude are relatively easy to control experimentally, we expect these results to be relevant to experiments in nonlinear optics and Bose-Einstein condensates, extending to coupled systems observations made for the scalar equation.
}

\section{Introduction}
\label{sec-intro}

In this paper we study the focusing vector nonlinear Schr\"odinger (NLS) equation, featuring quadratic dispersion and cubic nonlinearity of the slowly varying amplitude of two quasimonochromatic wave fields \cite{Man74},
\begin{equation}
\begin{split}
i \epsilon \pp{\psi_1}{t} + \frac{\epsilon^2}{2} \frac{\partial^2 \psi_1}{\partial x^2} + ( |\psi_1|^2 + |\psi_2|^2) \psi_1=&0 \\
i \epsilon \pp{\psi_2}{t} + \frac{\epsilon^2}{2} \frac{\partial^2 \psi_2}{\partial x^2} + ( |\psi_1|^2 + |\psi_2|^2) \psi_2=&0 \\
\end{split}
\label{eq-Man-NLS}
\end{equation}
where $\psi_1$ and  $\psi_2$ are the complex wave envelope, and $\epsilon$ is the dispersion parameter defining the modulation scale, with the initial condition
\begin{subequations}\label{eq:vec-init}
\begin{equation}\label{eq:vec-init-a}
\psi_k(x,0)=\sqrt{\mu_k(0)+\gamma_k(0)\,x^2}\,\exp\left(\frac{i\,\alpha_k(0)}{2\,\epsilon}\,x^2\right)\,,\qquad\mbox{with}\quad \mu_k(0)>0\,,\quad\gamma_k(0)<0\,,\qquad k=1,2\,,
\end{equation}
defined on the compact support
\begin{equation}\label{eq:vec-init-b}
x\in\left[-\sqrt{-\frac{\mu_k(0)}{\gamma_k(0)}},\sqrt{-\frac{\mu_k(0)}{\gamma_k(0)}}\right]\,,\qquad k=1,2\,.
\end{equation}
\end{subequations}
The end points of this compact support are vacuum points, where the initial condition features a catastrophe of the gradient. \colorblue{In each component, $\mu_{k}(0)$ fixes the peak intensity of the initial profile and $\gamma_{k}(0)$ its curvature, the two together fixing the width of the support through (\ref{eq:vec-init-b}). The quadratic phase in (\ref{eq:vec-init-a}) is the initial chirp, the quantity most readily controlled in an experiment.} In what follows we refer to system (\ref{eq-Man-NLS}) as the vector NLS equation.

The vector NLS equation was introduced by Manakov \cite{Man74} in the context of the self-focusing of electromagnetic waves of arbitrary polarisation, where it was shown to be integrable via the inverse scattering transform and its soliton solutions were constructed. Zakharov and Schulman \cite{ZS82} subsequently established that, up to rescaling, only three choices of the relative sign between the self- and cross-focusing coefficients preserve integrability, namely the defocusing, the mixed, and the focusing regime considered in this paper. Further families of exact soliton solutions, including the energy-redistributing collisions peculiar to the vector case, were obtained within the same inverse scattering framework \cite{RL95,RLH97,AKNS74}. 
In this paper we refer to system (\ref{eq-Man-NLS}) as the vector NLS equation, but in the literature it is also commonly referred to as the Manakov system or the coupled NLS system.

This work aims at extending
to the two-component case the findings established for the scalar case, 
for parabolic initial conditions on a compact support. For the scalar focusing NLS equation,
\begin{equation}\label{eq-Scalar-NLS}
i\epsilon\psi_t+\frac{\epsilon^2}{2}\psi_{xx}+|\psi|^2\psi=0\,,
\end{equation}
this Cauchy problem, with initial condition
\begin{subequations}
\begin{equation}
\psi(x,0)=\sqrt{\mu_0+\gamma_0\,x^2}\,\exp\left(\frac{i\,\alpha_0}{2\,\epsilon}\,x^2\right)\,,\qquad\mbox{with}\quad\mu_0>0\,,\quad\gamma_0<0\,,
\end{equation}
defined on the compact support
\begin{equation}
x\in\left[-\sqrt{-\frac{\mu_0}{\gamma_0}},\sqrt{-\frac{\mu_0}{\gamma_0}}\right]\,,
\end{equation}
\end{subequations}
was first posed and solved, for generic chirp $\alpha_0$, by Talanov in his famous work \cite{Tal65}, in the context of the self-focusing of light beams in cubic (Kerr) media. In what follows we refer to this single-field system as the scalar NLS equation. By Talanov dynamics we mean, throughout this paper, the evolution of a compactly supported, concave parabolic-profile initial condition satisfying, at $t=0$, the Talanov ansatz, for an equation -- such as (\ref{eq-Man-NLS}) or (\ref{eq-Scalar-NLS}) -- whose dispersionless limit admits such an ansatz.

Similarly to the scalar case, the vector NLS system is also deemed a universal model, as it emerges via multiscale reduction from a broad range of different, generically non-integrable, physical contexts. It was derived, prior to the proof of its integrability, as the natural envelope equation for two circularly polarised waves in an isotropic Kerr medium \cite{BZak70}. It arises
as the effective envelope system of the physically realistic, non-integrable propagation equations of a birefringent optical fibre \cite{WMC91,Men99}, and it emerges from the coupled Gross-Pitaevskii equations of a two-component Bose-Einstein condensate with matched scattering lengths \cite{KNFMC04}. \colorblue{More recently, a coupled-NLS-type system has been derived from a fluid-Maxwell model for two electromagnetic wavepackets of opposite circular polarisation co-propagating in a magnetised plasma \cite{VLFK24}.}
Beyond these derivations, the vector NLS system finds direct application across a wide variety of further contexts: crossing sea states in fluid dynamics \cite{OOS06,CBTBSO12,TBOSMO11,OPT10,AH15,ORBMA13}, multi-component Bose-Einstein condensates \cite{KU12,SKU13,KF16,PS02,PSt16,BKA09}, polarisation dynamics in birefringent nonlinear optical fibres \cite{Agr,HK95,Men87,MM06,MMW97,EMGB92}, and the nonlinear coupling of electrostatic wavepackets in plasma physics \cite{SGD79,RS97,SE07,MB89}. \colorblue{The magnetised, birefringent ionospheric plasma supports two co-propagating modes governed by a coupled NLS system \cite{BZak70,VLFK24}, and the initial data (\ref{eq:vec-init}) may accordingly be regarded as describing a pair of parabolic radio beams interacting with the medium and with each other.}

For the scalar case, the same Cauchy problem was first studied, in the dispersionless (geometric optics) limit, \colorblue{precisely }in the context of the interaction of parabolic radio beams with the ionosphere, by Talanov, Gurevich and Shvartsburg \cite{Tal65,GS73}, who obtained the blow-up criterion and catastrophe time for a subclass of the initial data. More recently, for zero chirp, the effect of modulational instability on the formation of a region of strong nonlinear oscillations was studied by Kamchatnov and Shaykin \cite{KS21}. This subclass of initial data was later considered in full generality in our previous work \cite{DORS23}. 
There, we completed the classification of the compactly supported self-similar solutions of the dispersionless scalar NLS equation for a generic concave parabolic profile and generic chirp. We numerically compared, in the semiclassical, small-dispersion regime, the dispersionless catastrophe criterion against the fully dispersive dynamics, showing that the latter is driven by the former. Whenever the dispersionless solution blows up in finite time the fully dispersive solution develops a high-order rogue wave whose onset time is accurately predicted by the dispersionless catastrophe time, and, conversely, the dispersive solution relaxes whenever its dispersionless counterpart does. This dynamics was understood, at the time, to interpolate between two known scenarios for rogue wave formation from localised initial data: the Bertola-Tovbis scenario \cite{BT10,BT13}, in which a smooth, bell-shaped wave packet undergoes a gradient catastrophe universally regularised by dispersion into a Peregrine-soliton-like structure, and the El-Khamis-Tovbis scenario \cite{ElKT16}, in which sharply discontinuous initial data give rise to counter-propagating dispersive dam-break flows that interact to form a modulated breather lattice. Our compactly supported, concave parabolic initial condition does not fit exactly into either scenario, yet we found evidence that, depending on the sign and magnitude of the chirp, the ensuing dynamics more closely resembles one or the other. 
The possible relevance of the Talanov self-similar solutions to the study of the fully dispersive focusing NLS equation in the case of small dispersion – to which our previous work \cite{DORS23} brought potential evidence in support – was pointed out by Suleimanov \cite{Sul17}, who also conjectured, for the zero-chirp case, a connection with the Painlevé-III hierarchy. Subsequently, this connection was rigorously established by Buckingham, Jenkins and Miller \cite{BJM25}, via a semiclassical soliton ensemble approach, for the broader class of so-called semicircular Klaus-Shaw potentials, of which the zero-chirp Talanov profile is a special case.
While the evolution of a uni-modal, compactly supported initial condition for the focusing scalar NLS equation, in the low-dispersion, semiclassical regime, has received much attention, \colorblue{}the equivalent problem for a system of two coupled NLS equations in the focusing regime has, to the best of our knowledge, escaped scrutiny, despite the vector NLS system being one of the most studied integrable systems in the literature. Vector rogue waves and vector solitons of the constant-coefficient, focusing vector NLS system have been extensively constructed via the inverse scattering transform and Darboux-dressing methods \cite{BDCW12,BCDLOW14,DLS19F,DLS18J,DL07,DL09,DL13,DL16,DLS19CRC,CBSGM17,ZBM15,LGZ14,ZL13,GL11}, their spectral and modulational stability has been analysed via squared eigenfunctions and standard Fourier methods \cite{FMMW00,LZ19}, \colorblue{and }their semiclassical modulational instability has been characterised both experimentally and theoretically \cite{FKFMBCM15}\colorblue{. Finally, the} underlying inverse scattering formalism has been generalised to the broader matrix NLS hierarchy, of which the vector NLS system is the column-vector reduction \cite{PDLH18,DPvdMV13,DPvdMV14,vdM15,PAB06,PBT11}. None of these approaches, however, addresses the dispersionless self-similar (Talanov) dynamics and the associated catastrophe criterion that are the subject of the present paper.

In analogy to what we have done in our previous work \cite{DORS23} for the scalar case, in this paper we provide, for arbitrary initial chirps, closed-form, implicit expressions for the evolution of compactly-supported, parabolic initial data of the form (\ref{eq:vec-init}) for the dispersionless vector NLS equation. In this setting, we derive a criterion for blow-up at finite time and an explicit expression for the catastrophe time whenever a gradient catastrophe occurs.
This construction describes the dispersionless dynamics faithfully only in a neighbourhood of the parabolic maxima, under the assumption that the two supports continue to overlap in that neighbourhood. We prove that this is guaranteed whenever the two fields share identical initial supports (and that the construction is in fact exact, rather than approximate, when the initial chirps coincide too). In this case, we further prove that catastrophe, when it occurs, always affects both fields simultaneously. The coupling alone can force a catastrophe even when both fields, if decoupled, would relax. The coupled catastrophe time is always earlier than the relevant decoupled catastrophe time, a property we verify extensively via direct numerical comparison. Finally, we prove that, for identical initial supports, in the semiclassical, small-dispersion regime, the dispersionless dynamics constructed here drives the evolution of the fully dispersive vector NLS equation, in direct analogy with the scalar case.

The paper is structured as follows. In Sec.~\ref{sec-MAnak} we introduce the dispersionless vector NLS system and its fluid-dynamical (Madelung) form. In Sec.~\ref{sec-TalMan} we construct the Talanov self-similar ansatz and solve the resulting Cauchy problem, separately discussing the branches $\alpha_1(0)=\alpha_2(0)$ and $\alpha_1(0)\neq\alpha_2(0)$. Section~\ref{sec-identical-supports} is devoted to the case of identical initial supports, for which we prove that the relative ordering of the two supports is preserved throughout the evolution and classify the effect of coupling on the catastrophe. In Sec.~\ref{sec-dispersive} we numerically compare the dispersionless dynamics with the fully dispersive one, for identical initial supports. Section~\ref{sec-conclusions} concludes the paper.

\section{Hydrodynamic formulation and the dispersionless system}
\label{sec-MAnak}
As in the scalar case, the vector NLS system~(\ref{eq-Man-NLS}) admits a fluid-dynamical analogue, obtained via an adapted version of the Madelung map \cite{Mad27}:
\begin{equation}
\psi_k=\sqrt{\rho_k(x,t)} \exp \left( \frac{i}{\epsilon}\int^x u_k(s,t)\, \D s \right)\,  , \qquad k=1,2\, .
\end{equation}
In terms of the densities $\rho_i$ and velocities $u_i$, the vector NLS system (\ref{eq-Man-NLS}) reads
\begin{equation}
\begin{split}
&\pp{\rho_i}{t}+ \pp{}{x} \left( \rho_i u_i\right)=0\, , \\
&\pp{u_i}{t}+ \pp{}{x} \left( \frac{1}{2}u_i^2 - \rho_1- \rho_2\right)
+ \epsilon^2 \pp{}{x} \left( \frac{1}{2 \sqrt{\rho_i}}  \frac{\partial^2}{\partial x^2}\sqrt{\rho_i} \right)=0\, , \qquad i=1,2\, .
\end{split}
\label{eq-Man-fluid}
\end{equation}
As in the classical scalar NLS case, the constant $\epsilon$ naturally introduces different space and time scales when $\epsilon \ll 1$.
The Madelung form (\ref{eq-Man-fluid}), unlike (\ref{eq-Man-NLS}), clearly separates the contributions of the fast and slow timescales.
The dispersionless vector NLS system is obtained from (\ref{eq-Man-NLS}) in the limit $\epsilon \to 0$:
\begin{equation}
\begin{pmatrix}
\rho_1 \\ u_1 \\ \rho_2 \\ u_2
\end{pmatrix}_t +
\begin{pmatrix}
u_1 & \rho_1 & 0 & 0 \\
-1 & u_1 & -1 & 0 \\
0 & 0 & u_2 & \rho_2 \\
-1 &0 & -1 & u_2
\end{pmatrix}
\begin{pmatrix}
\rho_1 \\ u_1 \\ \rho_2 \\ u_2
\end{pmatrix}_x=
\begin{pmatrix}
0\\ 0\\ 0 \\0
\end{pmatrix}\, .
\label{eq-dMan}
\end{equation}
\begin{remark}
The Haantjes tensor of the velocity matrix vanishes, and therefore, by \cite{FM07}, four Riemann invariants exist \colorblue{wherever the velocity matrix is diagonalisable}. We note that this does not hold for generic coupling constants.
\end{remark}
To establish whether the dispersionless vector NLS system (\ref{eq-dMan}) is hyperbolic or elliptic, we compute the characteristic polynomial of the velocity matrix $V$:
\begin{equation}
\begin{split}
\Pi_\lambda=\lambda ^4&-2 \left( u_1+u_2\right) \lambda ^3+ \left(\rho _1+\rho _2+u_1^2+4 u_2 u_1+u_2^2\right) \lambda ^2+ \\
&-2 \left(\rho _2 u_1+ \rho _1 u_2+ u_2 u_1^2+u_2^2 u_1\right)\lambda +\rho _1 u_2^2+\rho _2 u_1^2+u_1^2 u_2^2\, .
\end{split}
\label{eq:char-poly}
\end{equation}
Equivalently, introducing the variables $u_1=v+w$, $u_2=v-w$ and $\lambda= \ell+v$, this can be rewritten in depressed form as
\begin{subequations}\label{eq:PiL}
\begin{equation}
\Pi_\ell=\zeta_{0}+\zeta_{1}\,\ell+\zeta_{2}\,\ell^2+\ell^4\, ,
\end{equation}
where
\begin{equation}
\zeta_{0} = w^2(w^2+\rho _1+\rho _2)\,,\qquad
\zeta_{1} = 2w\left( \rho _1- \rho _2\right)\,,\qquad
\zeta_{2} = -2 w^2+\rho _1+\rho _2\,.
\end{equation}
\end{subequations}
The system is hyperbolic if these characteristic velocities -- the roots of $\Pi_\lambda$, equivalently of $\Pi_\ell$ -- are all real, and elliptic if they are all non-real.
\begin{prop}
\label{prop:ellipticity}
For every $\rho_1,\rho_2>0$ and every real \colorblue{$u_1\neq u_2$}, the characteristic polynomial $\Pi_\lambda$ in (\ref{eq:char-poly}) has no real roots: the dispersionless vector NLS system (\ref{eq-dMan}) is purely elliptic.
\end{prop}
\begin{proof}
\colorblue{The characteristic polynomial (\ref{eq:char-poly}) factorises as}
\begin{equation*}
\Pi_\lambda=\left(\lambda-u_1\right)^2\left(\lambda-u_2\right)^2+\rho_1\left(\lambda-u_2\right)^2+\rho_2\left(\lambda-u_1\right)^2\, .
\end{equation*}
\colorblue{For real $\lambda$, each term is non-negative, and they cannot vanish simultaneously when $u_1\neq u_2$, since the second requires $\lambda=u_2$ and the third $\lambda=u_1$. Hence $\Pi_\lambda>0$ for every real $\lambda$.}
\end{proof}
\begin{remark}
In a vacuum point of one of the two fields -- say, $\rho_1=0$ -- the velocity matrix becomes
\begin{equation}
V|_{\rho_1=0}=
\begin{pmatrix}
u_1 & 0 & 0 & 0 \\
-1 & u_1 & -1 & 0 \\
0 & 0 & u_2 & \rho_2 \\
-1 &0 & -1 & u_2
\end{pmatrix}
\end{equation}
with eigenvalues $u_1,u_1,u_2\pm i \sqrt{\rho_2}$. Since the matrix $V|_{\rho_1=0} - u_1 \mathbb{I}_4$ has corank $1$ while the eigenvalue $u_1$ has algebraic multiplicity $2$, the velocity matrix $V|_{\rho_1=0}$ is non-diagonalisable. Therefore, the system is parabolic in the vacua of at least one field.
\end{remark}

\colorblue{\begin{remark}
On the locus $u_1=u_2=u$ the factorisation in the proof of Proposition \ref{prop:ellipticity} reduces to
\begin{equation*}
\Pi_\lambda=\left(\lambda-u\right)^2\left[\left(\lambda-u\right)^2+\rho_1+\rho_2\right]\, ,
\end{equation*}
so that $\lambda=u$ is a real root of algebraic multiplicity two, while the remaining pair $\lambda=u\pm i\sqrt{\rho_1+\rho_2}$ is non-real. Since $V-u\,\mathbb{I}_4$ has corank $1$, $V$ is not diagonalisable, and the system degenerates to parabolic type on $u_1=u_2$, exactly as it does at the vacuum points.
\end{remark}}

\subsection{Conserved quantities of the dispersionless system}
System (\ref{eq-dMan}) admits the following conserved quantities, given as spatial integrals of the corresponding densities:
\begin{subequations}\label{eq:conserved-quantities}
\begin{align}
M_i &= \int_{-\infty}^{+\infty} \rho_i \, \mathrm{d}x\, , \qquad i=1,2\, , \\
P &= \int_{-\infty}^{+\infty} m \, \mathrm{d}x\, , \qquad m=\rho_1 u_1+\rho_2 u_2\, , \\
H &= \int_{-\infty}^{+\infty} h \, \mathrm{d}x\, , \qquad h=\frac{1}{2}\rho_1 u_1^2+\frac{1}{2}\rho_2 u_2^2-\frac{1}{2}(\rho_1+\rho_2)^2\, ,
\end{align}
\end{subequations}
namely the two masses $M_1,M_2$, the total momentum $P$, and the energy $H$, with respective densities $\rho_i$, $m$ and $h$.
Mathematically, the quantities $u_i$ are also conserved densities, although they do not appear to play any useful role in the model.
The dispersionless vector NLS system is Hamiltonian with respect to the canonical structure and the Hamiltonian $H$:
\begin{equation}
\begin{pmatrix}
\rho_1 \\ u_1 \\ \rho_2 \\ u_2
\end{pmatrix}_t +
\begin{pmatrix}
0 & \partial_x &0 &0 \\
 \partial_x &0 & 0 &0 \\
0 & 0 &0 &\partial_x \\
0 &0 &\partial_x &0
\end{pmatrix}
\begin{pmatrix}
\partial h/ \partial \rho_1 \\ \partial h/ \partial u_1 \\ \partial h/ \partial \rho_2 \\ \partial h/ \partial u_2
\end{pmatrix}=
\begin{pmatrix}
0\\ 0\\ 0 \\0
\end{pmatrix}\, .
\end{equation}
\begin{remark}
It is natural to ask whether the total momentum $P$ can play the role of an alternative Hamiltonian. The same equations can indeed be written by means of an antisymmetric tensor:
\begin{equation}
\begin{pmatrix}
\rho_1 \\ u_1 \\ \rho_2 \\ u_2
\end{pmatrix}_t +
\frac{1}{2}
\begin{pmatrix}
\rho_1 \partial_x + \partial_x \rho_1 & u_1 \partial_x &0 &0 \\
\partial_x u_1 & -2 \partial_x & 0 &-2\partial_x \\
0 & 0 & \rho_2 \partial_x + \partial_x \rho_2 & u_2 \partial_x \\
0 &-2\partial_x &\partial_x u_2 & -2 \partial_x
\end{pmatrix}
\begin{pmatrix}
\partial m/ \partial \rho_1 \\ \partial m/ \partial u_1 \\ \partial m/ \partial \rho_2 \\ \partial m/ \partial u_2
\end{pmatrix}=
\begin{pmatrix}
0\\ 0\\ 0 \\0
\end{pmatrix}\, .
\end{equation}
This tensor, however, is not Poisson, since the associated Dubrovin-Novikov metric is not flat (see \cite{DN83}).
An open question is whether the dispersionless vector NLS system provides an example of an infinite-dimensional generalisation of Poisson quasi-Nijenhuis type \cite{FMOP20, EFMP25}.
\end{remark}
A conserved density $q(\rho_1,\rho_2,u_1,u_2)$ satisfies the relation $q_t+J_x=0$ for a suitable current $J=J(\rho_1,\rho_2,u_1,u_2)$ such that
\begin{subequations}
\begin{align}
J_{\rho_1}&= u_1\,q_{\rho_1}-q_{u_1}-q_{u_2}\,,&
J_{u_1}&= \rho_1\,q_{\rho_1}+u_1q_{u_1}\,, \\
J_{\rho_2}&= u_2\,q_{\rho_2}-q_{u_1}-q_{u_2}\,,&
J_{u_2}&= \rho_2\,q_{\rho_2}+u_2 q_{u_2}\,.
\end{align}
\end{subequations}
The compatibility conditions for $J$ fix a system of PDEs for the conserved density $q$:
\begin{subequations}\label{concondM}
\begin{align}
& q_{\rho_1 u_1}+\left(u_1-u_2\right)\,q_{\rho_1 \rho_2}+q_{\rho_1 u_2}-q_{u_1 \rho_2}-q_{\rho_2 u_2}=0\, ,\\
& \rho_1\,q_{\rho_1 \rho_1}+q_{u_1 u_1}+q_{u_1 u_2}=0\, ,\\
& \rho_2\,q_{\rho_1 \rho_2}-\left(u_1-u_2\right)\,q_{\rho_1 u_2}+q_{u_1 u_2}+q_{u_2 u_2} =0\, ,\\
&  \rho_1\,q_{\rho_1 \rho_2}+\left(u_1-u_2\right)\,q_{u_1 \rho_2}+q_{u_1 u_1}+q_{u_1 u_2} =0\, ,\\
& q_{u_1 u_2}+ \rho_2\,q_{\rho_2 \rho_2}+q_{u_2 u_2} =0\, ,\\
&  \rho_1\,q_{\rho_1 u_2}- \rho_2\,q_{u_1 \rho_2}+\left(u_1-u_2\right)\,q_{u_1 u_2}=0\, .
\end{align}
\end{subequations}
The structure of system (\ref{concondM}) suggests the existence of a possibly infinite class of conserved densities sharing two properties: they are symmetric under the exchange
$(\rho_1,u_1) \leftrightarrow (\rho_2,u_2)$, and they are homogeneous with respect to the weights $\rho_i \to \tau^2 \rho_i$ and $u_i \to \tau u_i$.
Using these two properties, one can consider the ansatz:
\begin{equation}
h_n= \rho_1 u_1^{n-2}+\rho_2 u_2^{n-2} + \sum_{k=0}^{n-3} \sum_{i=0}^{\lfloor k/2 \rfloor}
C_{i,k-i}(\rho_1,\rho_2) \left(u_1^i u_2^{k-i}+ u_1^{k-i} u_2^i\right)\, ,
\end{equation}
where $C_{i,j}(\rho_1,\rho_2)$ are symmetric homogeneous polynomials of degree $i+j$.
It is possible to find an infinite number of conserved densities $\{h_n\}_{n=3,4,5,\dots}$, homogeneous in $\tau$, whose first few elements are:
\begin{subequations}\label{dMan-cc}
\begin{align}
h_3&=m=\rho_1 u_1+\rho_2 u_2\, ,\\
h_4&=2h= \rho_1 u_1^2+\rho_2 u_2^2-\left(\rho_1+\rho_2\right)^2\, ,\\
h_5&=\rho_1\,u_1^3 +\rho_2\,u_2^3 -3\,\left(\rho_1^2+ \rho_1\,\rho_2\right)\,u_1-3\,\left(\rho_2^2+ \rho_1\,\rho_2\right)\,u_2 \, , \\
h_6&= {\rho_1\,u_1^4}+{\rho_2\,u_2^4}-\left(6\,\rho_1^2+4\,\rho_1\,\rho_2\right)\,u_1^2-4\,\rho_1\,\rho_2\,u_1\,u_2-\left(6\,\rho_2^2+4\,\rho_1\,\rho_2\right)\,u_2^2+2\,\left(\rho_1+\rho_2\right)^3\, .
\end{align}
\end{subequations}

\section{Self-similar solutions and Talanov dynamics}
\label{sec-TalMan}
Following Barenblatt and Zel’dovich\footnote{If necessary we recall here the phrase used in \cite{BZ72} to stress the relevance of the self-similar solutions: \emph{``Thus, in practice self-similarity is of interest not because it, as it were, facilitates the calculation, but because it reflects an important property of the process at hand, which is obscured by accidental details of the initial and boundary conditions''}} in \cite{BZ72}, we look at the self-similar solutions of (\ref{eq-dMan}) to understand the intermediate asymptotics of the fully dispersive system (\ref{eq-Man-fluid}). Self-similar solutions of the first kind are obtained by dimensional analysis of the model.
In systems of conservation laws, typically the only parameter in play has the dimensions of a velocity, and the self-similar variable is taken to be $\xi=x/t$.
However, in the vector NLS case (as in the defocusing NLS), such a choice leads to the solutions
\begin{equation}
u_1=u_2=\frac{2}{3} \xi\, , \qquad \rho_1=-\frac{\cos^2(\theta)}{9} \xi^2\, , \qquad \rho_2=-\frac{ \sin^2(\theta)}{9} \xi^2\, \qquad \theta \in [0,2\pi)\, .
\end{equation}
Since $\rho_i$ must be positive by construction, these solutions are unphysical: this is not surprising, since the vector NLS system is elliptic.
However, one can relax the hypothesis that the structure of the solutions is fixed by dimensional analysis, and look instead for self-similar solutions of the second kind, where a generic field $f$ has the behaviour $f = F(t)\,G(\xi)$ with $\xi=x/t^\lambda$.
In this case, the solutions are obtained with the choices:
 \begin{equation}
\xi=\frac{x}{d}\, , \qquad d(t)=t^{2/3}\, ,
\end{equation}
which gives the same chirp as in the previous case, but physically relevant amplitudes:
\begin{equation}
\begin{split}
u_1&=u_2=\frac{2}{3 d(t)^{1/2} }  \xi = \frac{2}{3t} x \, ,  \\
\rho_1&= \frac{1}{d(t)}  \left( -\frac{\cos^2(\theta)}{9} \xi^2 + r_1 \right)=  -\frac{\cos^2(\theta)}{9 t^2} x^2  + \frac{r_1}{t^{2/3}}\, ,  \\
\rho_2&= \frac{1}{d(t)}  \left( -\frac{\sin^2(\theta)}{9} \xi^2 + r_2 \right) =  -\frac{\sin^2(\theta)}{9 t^2} x^2  + \frac{r_2}{t^{2/3}} \, ,
\end{split}
\end{equation}
where $\theta, r_1, r_2$ are arbitrary constants.
In his
 paper \cite{Tal65}, Talanov noted that, for the dispersionless focusing NLS, the class of self-similar solutions of the second kind
is much larger than expected. Following the same line of reasoning, and using the gradation of the conserved densities (\ref{dMan-cc}), we use the ansatz
\begin{equation}
\rho_i=\gamma_i(t) x^2+\mu_i(t)\, , \qquad u_i=\alpha_i(t) x\, \qquad i=1,2\, ,
\label{eq:TalMan}
\end{equation}
obtaining the following  system of ODEs for the variables $\alpha_i$, $\gamma_i$ and $\mu_i$:
\begin{subequations}\label{eq:ODEs-alpha-gamma-mu}
\begin{align}
&\dot{\alpha}_1+ \alpha_1^2 -2\,\left(\gamma_1+\gamma_2\right)=0\,,& &\dot{\alpha}_2+ \alpha_2^2 -2\,\left(\gamma_1+\gamma_2\right)=0\,,\label{eq:ODEs-alpha}\\
&\dot{\gamma}_1+3\,\alpha_1\,\gamma_1=0\,, & &\dot{\gamma}_2+3\,\alpha_2\,\gamma_2=0\,,\label{eq:ODEs-gamma}\\
&\dot{\mu}_1+ \alpha_1\,\mu_1=0\,,& &\dot{\mu}_2+ \alpha_2\,\mu_2=0\,,\label{eq:ODEs-mu}
\end{align}
\end{subequations}
Here and in what follows, a dot over a function denotes its derivative with respect to $t$.
\begin{remark}
As in the scalar case \cite{DORS23}, the ansatz (\ref{eq:TalMan}) is the only truncation of the general expansions $\rho_i=\sum_{k=0}^\infty \rho_{i,k}(t)\,x^k$ and $u_i=\sum_{k=0}^\infty u_{i,k}(t)\,x^k$, $i=1,2$, leading to an exact solution of the dispersionless vector NLS system (\ref{eq-dMan}). Indeed, inserting these expansions into (\ref{eq-dMan}) and collecting powers of $x$ shows that the nonlinear term $(\rho_i u_i)_x$ generates, at each order, only combinations of coefficients already present at that same order once $\rho_i$ is taken even and $u_i$ odd about the truncation degree, exactly as in the scalar case. Crucially, the coupling between the two fields enters only additively, through the sum $\rho_1+\rho_2$ in the equation for each $u_i$, and therefore does not raise the polynomial degree beyond what the scalar hierarchy already generates: the same degree bookkeeping that singles out $\rho=\gamma(t)x^2+\mu(t)$, $u=\alpha(t)x$ as the unique closing truncation for a single field continues to single out (\ref{eq:TalMan}) as the unique closing truncation for the coupled system, order by order in $\rho_1,\rho_2,u_1,u_2$ separately.
\end{remark}
Following the definition given in the Introduction, throughout the rest of this paper we refer to the evolution generated by the ansatz (\ref{eq:TalMan}) as Talanov dynamics whenever the initial data $\mu_i(0)$, $\gamma_i(0)$, $\alpha_i(0)$ are such that the resulting density profiles $\rho_i(x,0)$ are concave, parabolic, and compactly supported, as in (\ref{eq:vec-init}).
\begin{remark}\label{rem:validity}
As in the dispersionless defocusing NLS analysed by Talanov \cite{Tal65}, the solutions (\ref{eq:TalMan}) for $\gamma_i(0) < 0$ are defined only where the $\rho_i$'s are positive.
In the vector NLS case, however, the situation is more delicate, because the supports of the two fields are, in general, different, and the solutions are exact only where both
$\rho_i$'s are nonzero: this is the case near the maxima of the solutions (\ref{eq:TalMan}), which is the case studied in this paper.
\end{remark}

Consistently, in the rest of this paper we will assume $\gamma_1(0),\gamma_2(0)<0$ and consider the dispersionless solutions defined on the compact support:
\begin{subequations}\label{eq:support}
\begin{equation}\label{eq:compact-rho}
\rho_i=\left\{
\begin{array}{lcr}
\gamma_i(t) x^2+\mu_i(t)
&\mbox{for}&x\in\left[-L_i(t),L_i(t)\right]\\
0&\mbox{else}&\,
\end{array}\right.
\end{equation}
where
\begin{equation}\label{eq:interval}
L_i(t)=\sqrt{-\frac{\mu_i(t)}{\gamma_i(t)}}\,,\qquad i=1,2\,.
\end{equation}
\end{subequations}
In the scalar problem, a solution of the above form can simply be obtained by truncating the parabola where it meets the horizontal axis, giving an exact, compactly supported function whose support evolves in time: the truncation reflects the absence of the field outside the interval where the solution is positive, and nothing in the equation for the single density needs correcting to account for it. On the other hand, in the vector case this is more delicate. Truncating each field at its own zeros still produces two compactly supported profiles, but wherever one field's support extends beyond the other's, the ansatz continues to impose an interaction -- through the coupling term $\rho_1+\rho_2$ in (\ref{eq-dMan}) -- with a field that ought to be entirely absent there. In that region, the dispersionless solution ceases to be valid, let alone to approximate the true dispersive dynamics in the semiclassical limit $\epsilon\to0$. We expect our analysis of the dispersionless dynamics to approximate the actual dispersionless behaviour -- and hence to be relevant for the fully dispersive dynamics -- only in a neighbourhood of the parabolic maxima, provided the compact supports of the two fields (defined as the intervals where $\rho_{1}$ and $\rho_{2}$ separately vanish) continue to overlap in that neighbourhood throughout the whole time evolution. Therefore, in general, tracking the relative position of the two supports throughout the evolution is essential. However, we show below (Proposition \ref{prop:no-crossing}) that identical initial supports guarantee the two supports never swap the order in which one contains the other and the endpoints never pass each other. Under these assumptions, the dispersionless dynamics studied here can be expected to offer, in the small dispersion, semiclassical limit, a meaningful prediction for the evolution of a neighbourhood of the parabolic maxima for the fully dispersive case.

The equations for the $\alpha_i$'s are of Riccati type, which suggests the classical map
\begin{equation}
\alpha_i=\frac{ \dot{\sigma}_i}{\sigma_i}\,\,,
\quad\mbox{with}\quad
\sigma_i(0)=1\, , \quad \dot{\sigma}_i(0)=\alpha_i(0)\, ,\quad i=1,2\,,
\label{eq:alpha-sigma-map}
\end{equation}
which, via (\ref{eq:ODEs-gamma}) and (\ref{eq:ODEs-mu}), implies the following relations for the fields $\mu_i$ and $\gamma_i$:
\begin{equation}
\mu_i=\frac{\mu_i(0)}{\sigma_i}\, , \qquad \gamma_i=\frac{\gamma_i(0)}{\sigma_i^3}\, .
\label{eq:gamma-mu-sigma-map}
\end{equation}
Substituting (\ref{eq:alpha-sigma-map}) and (\ref{eq:gamma-mu-sigma-map}) into (\ref{eq:ODEs-alpha}), we have that the auxiliary fields $\sigma_i$ satisfy the following system of ODEs:
\begin{subequations}\label{eq:sigma-system}
\begin{equation}\label{eq:sigma-system-equations}
\begin{split}
&\ddot{\sigma}_1-2\sigma_1 \left( \frac{\gamma_1(0)}{\sigma_1^3} + \frac{\gamma_2(0)}{\sigma_2^3}\right)=0\, ,\\
&\ddot{\sigma}_2-2\sigma_2 \left( \frac{\gamma_1(0)}{\sigma_1^3} + \frac{\gamma_2(0)}{\sigma_2^3}\right)=0\, ,\\
\end{split}
\end{equation}
with initial condition
\begin{equation}
\sigma_i(0)=1\, , \qquad \dot{\sigma}_i(0)=\alpha_i(0)\, ,\qquad i=1,2\, .\label{eq:sigma-system-initialcondition}
\end{equation}
\end{subequations}
\begin{remark}
For the hyperbolic case, in the context of the shallow-water model, this mapping has been used in \cite{CFOPT19,CFOPP19}.
\end{remark}
In the next sections, we separately discuss the two cases $\alpha_{1}(0)=\alpha_{2}(0)$ and $\alpha_{1}(0)\neq\alpha_{2}(0)$.

\subsection{The case of identical initial chirps ($\alpha_1(0)=\alpha_2(0)$)}
\label{sec-alpha-equal}
Denote by $\alpha_0$ the common value of $\alpha_1(0)$ and $\alpha_2(0)$, $\alpha_1(0)=\alpha_2(0)=\alpha_{0}$. Let $\sigma(t)$ be the solution of the scalar Cauchy problem:
\begin{subequations}\label{eq:sigma-equal-ode}
\begin{align}
&\ddot\sigma-\frac{2\gamma_0}{\sigma^2}=0\,,\label{eq:sigma-equal-ode-a}\\
&\sigma(0)=1\,, \quad \dot\sigma(0)=\alpha_{1}(0)=\alpha_{2}(0)=\alpha_0\,.\label{eq:sigma-equal-ode-b}
\end{align}
\end{subequations}
It is immediate to see that setting $\sigma_1=\sigma_2=\sigma$ transforms either equation in (\ref{eq:sigma-system}) into the equation (\ref{eq:sigma-equal-ode-a}) satisfied by $\sigma$. In other words, the pair $(\sigma,\sigma)$ satisfies exactly the original coupled system (\ref{eq:sigma-system}), matching its initial data $\sigma_1(0)=\sigma_2(0)=1$, $\dot\sigma_1(0)=\dot\sigma_2(0)=\alpha_0$. Since (\ref{eq:sigma-system}) admits a unique solution for this data -- its right-hand side is smooth away from $\sigma_1=0$ or $\sigma_2=0$ -- then $(\sigma,\sigma)$ must be its unique solution. Therefore, $\sigma_1\equiv\sigma_2\equiv\sigma$ throughout the evolution, and hence, via (\ref{eq:alpha-sigma-map}), $\alpha_1(t)=\alpha_2(t)$ for all $t\geq0$. In particular, if the two initial supports coincide, $L_1(0)=L_2(0)$, then, for $\alpha_{1}(0)=\alpha_{2}(0)$, it is immediate to see from (\ref{eq:gamma-mu-sigma-map}) that $L_1(t)=L_{1}(0)\,\sigma(t)=L_{2}(0)\,\sigma(t)=L_{2}(t)$ for all $t$ throughout the solution's domain of existence.
\begin{remark}
\label{rem:exact-identical}
When, in addition to the initial chirps, the two initial supports coincide, $L_1(0)=L_2(0)$, the two supports remain equal throughout the evolution, $L_1(t)=L_2(t)$ for all $t$, as just shown. Consequently, as per the discussion following Remark \ref{rem:validity}, when both the initial chirps and the initial supports coincide, our construction with truncation (\ref{eq:compact-rho}) is an exact solution of the dispersionless system (\ref{eq-dMan}) for compactly-supported, parabolic initial data.
\end{remark}

Multiplying (\ref{eq:sigma-equal-ode}) by $\dot\sigma$ and integrating once, using the initial conditions (\ref{eq:sigma-equal-ode-b}), gives the first integral:
\begin{subequations}\label{eq:sigma-equal-first-integral}
\begin{equation}\label{eq:sigma-equal-first-integral-a}
\dot\sigma^2=A-\frac{4\gamma_0}{\sigma}\,,
\end{equation}
where
\begin{equation}\label{eq:sigma-equal-first-integral-b}
\gamma_0=\gamma_1(0)+\gamma_2(0)\,,\quad\mbox{and}\quad
A=\alpha_0^2+4\gamma_0\,.
\end{equation}
\end{subequations}
It is interesting to observe that $A$ features the same combination already singled out in the scalar case \cite{DORS23}, with the curvature of the single parabolic initial condition for the scalar problem replaced by the sum of the curvatures $\gamma_1(0)+\gamma_2(0)$ of the initial parabolic profiles in the two fields for the vector problem. By (\ref{eq:alpha-sigma-map}) and (\ref{eq:gamma-mu-sigma-map}) via (\ref{eq:sigma-equal-first-integral}), we see that the solution develops a gradient catastrophe exactly when $\sigma(t)$ reaches zero in finite time. Separating variables in (\ref{eq:sigma-equal-first-integral-a}), evaluating the resulting quadrature (\textit{e.g.}, see 2.228.1 in \cite{GR07}), and fixing the integration constant through $\sigma(0)=1$, yields the solution in implicit form.
\begin{prop}
\label{prop:alpha-equal}
Let $\gamma_1(0),\gamma_2(0)<0$ and $\alpha_1(0)=\alpha_2(0)=\alpha_0$. With $\gamma_0=\gamma_1(0)+\gamma_2(0)$ and $A=\alpha_0^2+4\gamma_0$, the solution of (\ref{eq:sigma-equal-ode}) can be expressed in implicit form as
\begin{subequations}\label{eq:sigma-equal-implicit}
\begin{equation}\label{eq:sigma-equal-implicit-a}
f(\sigma)^{2}=A^{3}(t-t_0)^2\,,
\end{equation}
with
\begin{equation}\label{eq:sigma-equal-implicit-b}
f(\sigma)=4\,\gamma_0\,\acoth\!\left(\frac{\sqrt{A}}{\sqrt{A-4\,\gamma_{0}/\sigma}}\right)+\sigma\,\sqrt{A}\,\sqrt{A-4\,\gamma_{0}/\sigma}\,,
\qquad\mbox{and}\qquad
t_0=-\frac{\alpha_0}{A}-\frac{4\,\gamma_0}{A^{3/2}}\,\atanh\!\left(\frac{\alpha_0}{\sqrt{A}}\right)\,,
\end{equation}
\end{subequations}
where $\acoth$ and $\atanh$ are understood, when their argument exceeds $1$ in modulus, by analytic continuation (equivalently, by the corresponding logarithmic expression), the imaginary parts cancelling between the two terms so that $f(\sigma)$ stays real. The time-dependent Talanov coefficients $\alpha_i$, $\gamma_i$, and $\mu_i$ then follow via (\ref{eq:alpha-sigma-map}) and (\ref{eq:gamma-mu-sigma-map}).
\end{prop}
To evaluate $\alpha_1=\alpha_2=\dot\sigma/\sigma$ in time, $\dot\sigma$ is needed explicitly as a function of $\sigma$ and $t$. Differentiating (\ref{eq:sigma-equal-implicit}) implicitly gives
\begin{equation}
\dot\sigma=\frac{\sqrt{A-4\,\gamma_0/\sigma}\,\left[\sqrt{A}\,(\alpha_0+A\,t)+4\,\gamma_0\,\atanh\!\left(\alpha_0/\sqrt{A}\right)\right]}{f(\sigma)}\, ,
\label{eq:sigma-dot-explicit}
\end{equation}
with $f(\sigma)$ as in (\ref{eq:sigma-equal-implicit-b}).
The solution develops a gradient catastrophe, at a finite time $t_c$, if and only if $\alpha_0<2\sqrt{-\gamma_0}$. Conversely, if $\alpha_0\geq2\sqrt{-\gamma_0}$, it undergoes asymptotic relaxation, with $\sigma(t)\to\infty$ as $t\to\infty$.
\begin{remark}
For $\alpha_0\neq-2\sqrt{-\gamma_0}$, the time of the gradient catastrophe $t_{c}$ follows from (\ref{eq:sigma-equal-implicit}) directly, evaluating in the limit $\sigma\to0^+$:
\begin{subequations}\label{eq:tcat-equal}
\begin{equation}
t_c=\frac{f(0)-f(1)}{A^{3/2}}=-\frac{\alpha_0+2\pi\sqrt{-\gamma_0}\,\sqrt{\gamma_0/A}}{A}-\frac{4\gamma_0}{A^{3/2}}\atanh\!\left(\frac{\alpha_0}{\sqrt A}\right)\,,
\quad\mbox{for}\quad
\alpha_0<2\sqrt{-\gamma_0}\,,\,\, \alpha_0\neq-2\sqrt{-\gamma_0}\,.
\label{eq:tcat-equal-general}
\end{equation}
For $\alpha_0=-2\sqrt{-\gamma_0}$ (i.e. $A=0$), (\ref{eq:tcat-equal-general}) is not directly applicable. The first integral (\ref{eq:sigma-equal-first-integral}) reduces instead to $\dot\sigma=-2\sqrt{-\gamma_0}/\sqrt\sigma$, integrating elementarily to $\sigma(t)=\left(1-3\sqrt{-\gamma_0}\,t\right)^{2/3}$. Therefore,
\begin{equation}
t_c=\frac{1}{3\sqrt{-\gamma_0}}\,,
\quad\mbox{for}\quad \alpha_0=-2\sqrt{-\gamma_0}\, ,
\label{eq:tcat-equal-degenerate}
\end{equation}
\end{subequations}
consistently with the scalar case \cite{DORS23} (where the symbol $\sigma$ denotes the reciprocal of our present $\sigma$). Note that $t_c$ does not depend on $\mu_1(0)$ or $\mu_2(0)$, and, like $\sigma(t)$ itself, depends on $\gamma_1(0)$, $\gamma_2(0)$ only through their sum $\gamma_0$: as in the scalar case, the time of the catastrophe is dictated entirely by the chirp and the total curvature, and any two pairs of curvatures sharing the same $\gamma_0$ give an identical $t_c$, even though $\gamma_1(t)=\gamma_1(0)/\sigma^3$ and $\gamma_2(t)=\gamma_2(0)/\sigma^3$ individually still differ from one another.
\end{remark}

\subsection{The case of different initial chirps ($\alpha_1(0)\neq\alpha_2(0)$)}
\label{sec-alpha-unequal}
We now assume $\alpha_1(0)\neq\alpha_2(0)$, and set
\begin{subequations}\label{eq:p-def-initialconditions}
\begin{equation}\label{eq:p-def}
p(t)=\frac{\sigma_1(t)}{\sigma_2(t)}\,,
\end{equation}
entailing, via (\ref{eq:sigma-system-initialcondition}),
\begin{equation}\label{eq:p-initialconditions}
p(0) = 1\,,\quad
\dot{p}(0) = \alpha_{1}(0)-\alpha_{2}(0)\,,\quad
\ddot{p}(0) = -2\,\alpha_{2}(0)\,\left[\alpha_{1}(0)-\alpha_{2}(0)\right]\,.
\end{equation}
\end{subequations}
Combining the two equations in (\ref{eq:sigma-system}) via $\sigma_1=p\,\sigma_2$ gives, for $\sigma_2$ and $p$, the pair
\begin{equation}
p^3\sigma_2^2\ddot\sigma_2=2\left(\gamma_1(0)+\gamma_2(0)p^3\right)\, , \qquad \ddot p\,\sigma_2+2\dot p\,\dot\sigma_2=0\, .
\label{eq:sigma2-p}
\end{equation}
The second equation is an exact differential of the form $\frac{\mathrm{d}}{\mathrm{d}t}\left(\dot p\,\sigma_2^2\right)=0$, yielding
\begin{equation}
\sigma_2=\frac{\sqrt{\alpha_1(0)-\alpha_2(0)}}{\sqrt{\dot p}}\, ,
\label{eq:sigma2-of-pdot}
\end{equation}
where the constant of integration was determined by imposing (\ref{eq:p-initialconditions}). Here and in the following, the half-integer powers of $\alpha_1(0)-\alpha_2(0)$ are understood by analytic continuation, being complex for $\alpha_1(0)<\alpha_2(0)$. We will ensure that the solution remains real during the whole time evolution. Substituting (\ref{eq:sigma2-of-pdot}) into the first equation of (\ref{eq:sigma2-p}) eliminates $\sigma_2$ altogether, giving the autonomous third-order ODE:
\begin{equation}
8\,\left[\gamma_1(0)+\gamma_2(0)\,p^3\right]\dot{p}^{\,7/2}-3\,\left[\alpha_1(0)-\alpha_2(0)\right]^{3/2}\,p^3\,\ddot{p}^{2}+
2\,\left[\alpha_1(0)-\alpha_2(0)\right]^{3/2}\,p^{3}\,\dot{p}\,\dddot{p}=0\,.
\label{eq:p-ode}
\end{equation}
We set $\dot{p}=y(p)$, so that $\ddot{p}=y\,y'$ and $\dddot{p}=y^{2}\,y''+y\,{y'}^{2}$, where primes denote differentiation with respect to $p$. Replacing in (\ref{eq:p-ode}) and using (\ref{eq:p-initialconditions}), we verify that $y$ satisfies the non-autonomous second-order ODE:
\begin{subequations}\label{eq:y-ode-initialcondition}
\begin{equation}\label{eq:y-ode}
8\,\left[\gamma_1(0)+\gamma_2(0)\,p^3\right]\,y^{3/2}-p^3\,\left[\alpha_1(0)-\alpha_2(0)\right]^{3/2}\,{y'}^2+
2\,p^{3}\,\left[\alpha_1(0)-\alpha_2(0)\right]^{3/2}\,y\,y''=0\,,
\end{equation}
satisfying the initial condition:
\begin{equation}\label{eq:y-initialcondition}
y(1)=\alpha_1(0)-\alpha_2(0)\,, \qquad y'(1)=-2\,\alpha_2(0)\,.
\end{equation}
\end{subequations}
Remarkably, equation (\ref{eq:y-ode}) can be written as an exact differential for $\sqrt{y}$:
\begin{equation}
\frac{\mathrm{d}^2\sqrt y}{\mathrm{d}p^2}=-\frac{2\left(\gamma_1(0)+\gamma_2(0)p^3\right)}{p^3\left(\alpha_1(0)-\alpha_2(0)\right)^{3/2}}\, ,
\label{eq:exact-diff}
\end{equation}
directly integrable twice in $p$. Carrying out the two integrations and fixing the integration constants via (\ref{eq:p-initialconditions}), we obtain that $p$ satisfies the first-order ODE:
\begin{subequations}\label{eq:p-final-coeffs}
\begin{equation}\label{eq:p-final}
\left[\alpha_1(0)-\alpha_2(0)\right]^{3}\,p^{2}\,\dot{p} =
\gamma_2(0)^2\,\left(c_0+c_1\,p+c_2\,p^2+p^3\right)^2\,,
\end{equation}
where
\begin{equation}\label{eq:cubic-coeffs}
\begin{split}
&c_0=\frac{\gamma_1(0)}{\gamma_2(0)}\,,\\
&c_1=\frac{-\alpha_1(0)^2+\alpha_1(0)\alpha_2(0)-2\gamma_1(0)+\gamma_2(0)}{\gamma_2(0)}\,,\\
&c_2=\frac{-\alpha_2(0)^2+\alpha_1(0)\alpha_2(0)-2\gamma_2(0)+\gamma_1(0)}{\gamma_2(0)}\,.\\
\end{split}
\end{equation}
\end{subequations}
Let $Q(p)$ be the monic polynomial featured on the right-hand side of (\ref{eq:p-final}) and let $p_1$, $p_2$, and $p_3$ be its roots:
\begin{equation}
Q(p)=c_0+c_1\,p+c_2\,p^2+p^3=(p-p_1)\,(p-p_2)\,(p-p_3)\,,
\label{eq:cubic}
\end{equation}
so that (\ref{eq:p-final}) can be rephrased as
$$
\left[\alpha_1(0)-\alpha_2(0)\right]^{3}\,p^{2}\,\dot{p} =\gamma_2(0)^2\,Q(p)^{2}\,.
$$
Since $Q(p)$ has real coefficients, $Q(p)^2\geq0$ for every real $p$. Then, (\ref{eq:p-final}) entails $\mathrm{sign}(\dot p)=\mathrm{sign}(\alpha_1(0)-\alpha_2(0))$ throughout the evolution (away from $p=0$). Thus, $p$ moves monotonically away from $p(0)=1$, decreasing towards $0$ if $\alpha_1(0)<\alpha_2(0)$, or increasing towards $+\infty$ if $\alpha_1(0)>\alpha_2(0)$. A further quadrature, carried out by partial fractions over $p_1$, $p_2$, $p_3$, then yields $p(t)$ implicitly.
\begin{prop}
\label{prop:alpha-unequal}
Let $\gamma_1(0),\gamma_2(0)<0$ and $\alpha_1(0)\neq\alpha_2(0)$. Let $c_0$ be as in (\ref{eq:cubic-coeffs}). Let $p_1$, $p_2$, and $p_3$ be the roots of $Q(p)$ as in (\ref{eq:cubic}). Then $p(t)=\sigma_1(t)/\sigma_2(t)$ is given implicitly by
\begin{subequations}\label{eq:p-sol}
\begin{equation}\label{eq:p-sol-implicit}
\hat{f}(p)=t-t_0\,,
\end{equation}
where
\begin{equation}\label{eq:fhat}
\hat{f}(p)=-\frac{\left[\alpha_1(0)-\alpha_2(0)\right]^3}{\gamma_2(0)^2}\sum_{\mathrm{cyc}}\left[\frac{2(c_0+p_1^3)}{(p_1-p_2)^3(p_1-p_3)^3}\log(p-p_1)+\frac{p_1^2}{(p_1-p_2)^2(p_1-p_3)^2\,(p-p_1)}\right]\,,
\end{equation}
with the cyclic sum running over $(p_1,p_2,p_3)\to(p_2,p_3,p_1)\to(p_3,p_1,p_2)$, and
\begin{equation}\label{eq:p-sol-t0}
t_0=-\hat{f}(1)\,.
\end{equation}
\end{subequations}
Here $\log$ is understood, where its argument is negative, by analytic continuation, consistently with (\ref{eq:sigma-equal-implicit}).
\noindent Correspondingly, using (\ref{eq:alpha-sigma-map}) and (\ref{eq:gamma-mu-sigma-map}) with (\ref{eq:sigma2-of-pdot}), we have
\begin{subequations}\label{eq:reconstruction-unequal}
\begin{align}
&\alpha_1=\frac{\gamma_2(0)^2\left(2\,c_0+c_1\,p-p^3\right)\,Q(p)}{\left[\alpha_1(0)-\alpha_2(0)\right]^3\,p^3}\,,&
&\alpha_2=\frac{\gamma_2(0)^2\left(c_0-c_2\,p^2-2\,p^3\right)\,Q(p)}{\left[\alpha_1(0)-\alpha_2(0)\right]^3\,p^3}\, ,\label{eq:reconstruction-unequal-a}\\
&\gamma_1=\frac{\gamma_1(0)\,|\gamma_2(0)|^3\,|Q(p)|^3}{\left[\alpha_1(0)-\alpha_2(0)\right]^6\,|p|^3\,p^3}\,,&
&\gamma_2=\frac{\gamma_2(0)\,|\gamma_2(0)|^3\,|Q(p)|^3}{\left[\alpha_1(0)-\alpha_2(0)\right]^6\,|p|^3}\, ,\label{eq:reconstruction-unequal-b}\\
&\mu_1=\frac{\mu_1(0)\,|\gamma_2(0)|\,|Q(p)|}{\left[\alpha_1(0)-\alpha_2(0)\right]^2\,|p|\,p}\,,&
&\mu_2=\frac{\mu_2(0)\,|\gamma_2(0)|\,|Q(p)|}{\left[\alpha_1(0)-\alpha_2(0)\right]^2\,|p|}\,.\label{eq:reconstruction-unequal-c}
\end{align}
\end{subequations}
\end{prop}
In practice, the Cauchy problem is solved as follows.
\begin{enumerate}
\item Compute $c_0,c_1,c_2$ from (\ref{eq:cubic-coeffs}) and the roots $p_1,p_2,p_3$ of (\ref{eq:cubic}).
\item Fix $t_0=-\hat f(1)$.
\item For each $t$, invert $\hat f(p)=t-t_0$ numerically for $p=p(t)$, continuing the root search in $t$ from $p(0)=1$, in the direction dictated by $\mathrm{sign}(\alpha_1(0)-\alpha_2(0))$.
\item Reconstruct $\alpha_1,\alpha_2,\gamma_1,\gamma_2,\mu_1,\mu_2$ via (\ref{eq:reconstruction-unequal}).
\end{enumerate}
By (\ref{eq:reconstruction-unequal}), $\alpha_{j}$, $\gamma_{j}$, $\mu_{j}$, $j=1,2$, if divergent, they all diverge at the same time $t_{c}$, and the solution features a catastrophe of the gradient, whenever $p$ approaches its limiting values, either $0$ or $+\infty$. At the catastrophe, the length of each support, $L_{i}$, $i=1,2$, contracts to zero. On the other hand, whenever $p(t)$ approaches one of the roots of $Q$, the system features asymptotic relaxation (and the length of each support diverges).
As for the evolution of the supports, we have the following fact.
\begin{prop}
\label{prop:no-crossing}
Let $\gamma_1(0),\gamma_2(0)<0$, and suppose the two initial supports coincide, $L_1(0)=L_2(0)$, with $\alpha_1(0)\neq\alpha_2(0)$. Let $T=t_{c}$ denote the time of the gradient catastrophe if one occurs, or $T=\infty$ if the solution instead relaxes. Then $L_1(t)\neq L_2(t)$ for every $t\in(0,T)$, and $L_1(t)-L_2(t)$ keeps, throughout this interval, the sign it takes immediately after $t=0$ -- the same sign as $\alpha_1(0)-\alpha_2(0)$. If $T<\infty$, the two supports coincide once more, and only then: $L_1(t),L_2(t)\to0$ together as $t\to T$.
\end{prop}
\begin{proof}
By (\ref{eq:gamma-mu-sigma-map}), $L_i(t)=\sigma_i(t)L_i(0)$. Since $L_1(0)=L_2(0)$, $L_1(t)=L_2(t)$ if and only if $\sigma_1(t)=\sigma_2(t)$, that is, if and only if $p(t)=1$. As already observed after equation (\ref{eq:cubic}), $\dot{p}$ vanishes only where $Q(p)$ does, and otherwise carries the fixed sign of $\alpha_1(0)-\alpha_2(0)$. Now $p(0)=1$, and directly from (\ref{eq:cubic-coeffs}),
$$
Q(1)=c_0+c_1+c_2+1=\frac{\left[\alpha_1(0)-\alpha_2(0)\right]^2}{|\gamma_2(0)|}>0\,,
$$
since $\alpha_1(0)\neq\alpha_2(0)$. The function $p$ moves away from $1$ immediately, in the direction fixed by $\mathrm{sign}(\alpha_1(0)-\alpha_2(0))$. For $p$ to return to $1$ within $(0,T)$, $\dot p$ would have to vanish somewhere along the way, which happens only where $p$ actually reaches a root of $Q$. However, in $(0,T)$ this never occurs. Indeed, near a simple root $p_{k}$ one has $Q(p)\approx Q'(p_{k})(p-p_{k})$, so
$$
\frac{\mathrm{d}t}{\mathrm{d}p}=\frac{\left[\alpha_1(0)-\alpha_2(0)\right]^3\,p^2}{\gamma_2(0)^2Q(p)^2}
$$
behaves like a nonzero constant divided by $(p-p_{k})^2$, and
$$
\int \frac{\mathrm{d}p}{(p-p_{k})^2}
$$
diverges as $p\to p_{k}$: reaching $p_{k}$ would take infinite time, entailing $T=+\infty$ (relaxation). If instead no root of $Q$ lies in the direction $p$ travels, there is nothing for $\dot{p}$ to vanish at in the first place, and the solution catastrophises at a finite $T=t_c$. Either way, $\dot{p}\neq0$ throughout $(0,T)$: $p$ is strictly monotonic there, never returning to the value $1$, and $L_1(t)-L_2(t)$ keeps a single fixed sign on $(0,T)$.
It remains only to identify what $L_1,L_2$ approach at $T$ itself, when $T<\infty$. If the solution catastrophises at a finite $T$, then $p\to0$ or $p\to\infty$ as $t\to T$. Using (\ref{eq:reconstruction-unequal}), it is immediate to see that:
\begin{align*}
L_{1}(t)^{2} &= -\frac{\mu_{1}(t)}{\gamma_{1}(t)} = -\frac{\mu_{1}(0)\,\left[\alpha_{1}(0)-\alpha_{2}(0)\right]^{4}\,|p|^{2}\,p^{2}}{\gamma_{1}(0)\,\left|\gamma_{2}(0)\right|^{2}\,\left|Q(p)\right|^{2}}\,,\\
L_{2}(t)^{2} &= -\frac{\mu_{2}(t)}{\gamma_{2}(t)} =
-\frac{\mu_{2}(0)\,\left[\alpha_{1}(0)-\alpha_{2}(0)\right]^{4}\,|p|^{2}}{\gamma_{2}(0)\,\left|\gamma_{2}(0)\right|^{2}\,\left|Q(p)\right|^{2}}\,.
\end{align*}
Thus, $L_{1}(t)$,$L_{2}(t)$ both approach zero as $p$ approaches zero or diverges.
\end{proof}
What this proposition shows is that, starting from two identical initial supports, the ordering set in the instant after $t=0$ is preserved for as long as the solution exists, whether it ends in relaxation or catastrophe. The key ingredient here is the coincidence of the initial supports: without that hypothesis, one can show that a reversal of which support contains the other becomes possible during the evolution. We leave a full treatment of that more general case, and of the resulting picture when the supports are not initially identical, to future work.
This is also what allows considering the dispersionless solution as informative about the fully dispersive evolution of the two density maxima. If the two initial supports coincide, then the two parabolic maxima in the two fields are defined in a spatial region in a neighbourhood of $x=0$ where the two supports continue to overlap for all times (possibly both contracting to vanishing measure at $x=0$ if the system features a catastrophe of the gradient), ensuring the validity and relevance of the $\rho_{1}$,$\rho_{2}$ interaction in (\ref{eq-dMan}) in that neighbourhood throughout the evolution. The comparison between the dispersionless case and the fully dispersive case for identical initial supports in the semiclassical regime will be carried out numerically in the last section this paper.
As for the gradient catastrophe, we can prove the following two facts, regardless of the ordering of $\alpha_1(0)$, $\alpha_2(0)$.
\begin{prop}
\label{prop:delta-sufficient}
Let $\gamma_1(0),\gamma_2(0)<0$ and
\begin{align*}
&D=\alpha_1(0)-\alpha_2(0)\,,\qquad S=\alpha_1(0)+\alpha_2(0)\,,\\
&g_{1} = \left[\alpha_1(0)-\alpha_2(0)\right]^{2} + \gamma_1(0)+\gamma_2(0)\,,\\
&g_{2} = \alpha_1(0)^2-\alpha_2(0)^2+3\,\left[\gamma_1(0)-\gamma_2(0)\right]\,.
\end{align*}
The discriminant of $Q$ with respect to $p$ is
\begin{subequations}\label{eq:discriminant}
\begin{equation}
\Delta_{p}(Q) = \frac{D^{2}}{48\,\gamma_{2}(0)^{4}}\,\Delta_{Q}\,,
\end{equation}
where
\begin{align}\label{eq:discriminant-DeltaQ}
\Delta_{Q}=192\,g_1^3+12\,g_1^2\,\left(S^2-36\,D^2\right)&+
36\,g_1\,D\,\left(9\,D^3+2\,g_2\,S-D\,S^2\right)+\notag\\
&-D\,\left[36\,g_2^2\,D-4\,g_2\,S\,\left(S^2-9\,D^2\right)+D\,\left(S^2-9\,D^2\right)^2\right]\,,
\end{align}
\end{subequations}
so that the sign of $\Delta_{p}(Q)$ coincides with the sign of $\Delta_{Q}$.
Then, if $\Delta_{Q}<0$, the solution undergoes a gradient catastrophe, whatever the ordering of $\alpha_1(0)$, $\alpha_2(0)$.
\end{prop}
\begin{proof}
From (\ref{eq:cubic-coeffs}), we have $Q(0)=c_0=\gamma_1(0)/\gamma_2(0)>0$. Since $Q$ is monic, $Q(p)\to-\infty$ as $p\to-\infty$. Hence, by the intermediate value theorem, $Q$ has a real root in $(-\infty,0)$. When $\Delta_{Q}<0$, $Q$ has exactly one real root. Combining with the previous observation, we have that, when $\Delta_{Q}<0$, $Q$ has one negative real root. A negative root lies in neither $(0,1)$ nor $(1,\infty)$, so -- whichever of these two intervals $p(t)$ actually travels monotonically through as $t$ evolves -- it does not reach any real root of $Q$. Thus, $p$ either approaches zero or diverges to $+\infty$, and does so in finite time. Indeed, from (\ref{eq:p-final}), the quantity
$$
\frac{1}{\dot{p}}=\frac{\left[\alpha_1(0)-\alpha_2(0)\right]^3\,p^2}{\gamma_2(0)^2\,Q(p)^2}
$$
stays bounded as $p\to0$ (since $Q(0)=c_0\neq0$) or decays like $p^{-4}$ as $p\to\infty$, so the corresponding quadrature converges at either endpoint.
\end{proof}
\begin{prop}
\label{prop:alpha-negative}
If $\alpha_1(0)<0$ or $\alpha_2(0)<0$, the solution undergoes a gradient catastrophe.
\end{prop}
\begin{proof}
As $p(t)$ evolves montonically with $p(t)>0$ for all $t$, from (\ref{eq:p-def}) and (\ref{eq:sigma2-of-pdot}) we have that $\sigma_j(t)>0$, $j=1,2$, for all $t$. Consequently, equation (\ref{eq:gamma-mu-sigma-map}) entails that $\gamma_j(t)<0$, $j=1,2$, throughout the evolution, so $2\,\left[\gamma_1(t)+\gamma_2(t)\right]<0$. Therefore, from (\ref{eq:alpha-sigma-map}), we have $\dot{\alpha}_i=2\,(\gamma_1+\gamma_2)-\alpha_i^2<0$, $i=1,2$, for all $t$. Therefore, $\alpha_i$, $i=1,2$, are strictly decreasing functions. Now, suppose the solution undergoes asymptotic relaxation, meaning $p(t)\to p_{k}$ as $t\to\infty$ for some root $p_{k}\neq0$ of $Q$ (the roots of $Q$ are always different from zero). By (\ref{eq:reconstruction-unequal}), $\alpha_1$ and $\alpha_2$ are each explicitly proportional to $Q(p)$, so $\alpha_{i}(t)\to0$, $i=1,2$, as $p\to p_{k}$. Conversely, if $\alpha_i(0)<0$ for some $i$, then since $\alpha_i(t)$ is strictly decreasing, $\alpha_i(t)<\alpha_i(0)<0$ for all $t>0$: $\alpha_i$ can only move further from $0$, never towards it, so $\alpha_i(t)\to0$ is impossible. If $\alpha_{i}(t)$ diverges for some $i$, then all the $\alpha$'s, $\gamma$'s and $\mu$'s diverge, entailing a gradient catastrophe.
\end{proof}
It remains to characterise the case $\gamma_1(0),\gamma_2(0)<0$, $\alpha_1(0)>0$, $\alpha_2(0)>0$, and $\Delta_{Q}\geq0$, the only one left open by Propositions \ref{prop:delta-sufficient} and \ref{prop:alpha-negative}. Here the two orderings, $\alpha_1(0)>\alpha_2(0)$ and $\alpha_1(0)<\alpha_2(0)$, genuinely differ, since $p$ travels towards different limits, and are treated in turn.
\begin{remark}
\label{rem:swap-symmetry}
The two cases, $\alpha_1(0)>\alpha_2(0)$ and $\alpha_1(0)<\alpha_2(0)$, can be mapped one into the other by exchanging $\alpha_1(0)\leftrightarrow\alpha_2(0)$, $\gamma_1(0)\leftrightarrow\gamma_2(0)$, and $p\leftrightarrow1/p$, the latter exchange being forced by the definition of $p$, (\ref{eq:p-def}). The initial values $\mu_1(0),\mu_2(0)$ play no role in either exchange.
\end{remark}

\subsubsection{The case $\alpha_1(0)<\alpha_2(0)$}
\label{sec-alpha-less}
If $\alpha_1(0)<\alpha_2(0)$, then $p$ decreases monotonically from $1$ towards $0$. Equation (\ref{eq:p-final}) shows that $\dot{p}$ is proportional to $Q(p)^2$, so $\dot{p}$ vanishes exactly where $Q$ does. Suppose $Q$ has a root $p_{k}$ inside $(0,1)$: as $p(t)$ decreases and approaches $p_{k}$, $\dot{p}$ approaches zero there too. If $\Delta_{Q}\neq0$, $p_{k}$ is a simple root of $Q$, and $\dot{p}$ vanishes quadratically, behaving like $(p-p_{k})^2$ near $p_{k}$. This is the same behaviour as a particle whose speed decreases in proportion to the square of its distance to a target: it gets arbitrarily close but never arrives in finite time. The same fact is visible directly in formula (\ref{eq:fhat}): the term $p_1^2/\bigl[(p_1-p_2)^2(p_1-p_3)^2(p-p_1)\bigr]$, and its two cyclic partners, each have a pole at $p=p_1$. If one of the roots, say $p_1$, equals $p_{k}$ and lies in $(0,1)$, then $\hat f(p)$ diverges as $p\to p_{k}$, and so does $t=\hat f(p)+t_0$. The solution takes infinitely long to approach $p_{k}$, and $p$ never reaches $0$: this is asymptotic relaxation. If, on the other hand, $Q$ has no root anywhere in $(0,1)$, then $Q(p)\neq0$ throughout the half-open interval $[0,1)$, $\dot p$ never vanishes there, and $\hat f(p)$ stays finite all the way down to $p=0$: the solution reaches $p=0$ at the finite time $t_c=\hat f(0)+t_0$, and this is the gradient catastrophe. In short: relaxation occurs exactly when $Q$ has a real root in $(0,1)$, and catastrophe occurs exactly when it does not.
\begin{lem}
\label{lem:root-in-01}
Let $\Delta_{Q}>0$, so that $Q$ has three distinct real roots, and let $p_{\max}<p_{\min}$ denote the locations of its local maximum and local minimum respectively. Then $Q$ has a real root in $(0,1)$ if and only if $0<p_{\min}<1$.
\end{lem}
\begin{proof}
Since $Q$ is monic, $Q(-\infty)=-\infty$ and $Q(+\infty)=+\infty$. With three real roots, $Q$ rises from $-\infty$ to a local maximum at $p_{\max}$, falls to a local minimum at $p_{\min}$, then rises to $+\infty$. As $Q(p_{\max})\,Q(p_{\min})=-\Delta_{p}(Q)/27<0$ for $\Delta_{Q}>0$, it must be $Q(p_{\max})>0$ and $Q(p_{\min})<0$.
Recall, from (\ref{eq:cubic-coeffs}) and (\ref{eq:cubic}), that $Q(0)=\gamma_1(0)/\gamma_2(0)>0$ and $Q(1)=(\alpha_1(0)-\alpha_2(0))^2/|\gamma_2(0)|>0$.
If $p_{\min}\leq0$, then, since $p_{\max}<p_{\min}$, one has $p_{\max}<0$, and $(0,1)$ lies entirely past both critical points, in the final increasing branch of $Q$. In this case, as $Q(0)>0$ and $Q$ increases throughout $(0,1)$, we have $Q(p)\geq Q(0)>0$, yielding no root in the interval $(0,1)$. If $p_{\min}\geq1$, then two configurations are possible for $p_{\max}$. If $p_{\max}\leq0$, then $(0,1)$ lies entirely in the decreasing branch between the two critical points. In this case, as $Q(1)>0$ and $Q$ decreases throughout $(0,1)$, we have $Q(p)\geq Q(1)>0$, yielding no root in the interval $(0,1)$. If instead $0<p_{\max}<1$, then $(0,1)$ splits at $p_{\max}$. $Q$ increases on $(0,p_{\max})$, so $Q(p)\geq Q(0)>0$ there and $Q$ decreases on $(p_{\max},1)$, so $Q(p)\geq Q(1)>0$ there. Either way, $Q$ stays positive throughout $(0,1)$, yielding no root in the interval $(0,1)$. On the other hand, if $0<p_{\min}<1$, then $Q(0)>0>Q(p_{\min})$, so by the intermediate value theorem $Q$ has a root in $(0,p_{\min})\subset(0,1)$.
\end{proof}
\begin{prop}
\label{prop:criterion-less}
Let $\gamma_1(0),\gamma_2(0)<0$, $\alpha_1(0)<\alpha_2(0)$, $\alpha_1(0)>0$, $\alpha_2(0)>0$, and $\Delta_{Q}>0$. The solution undergoes asymptotic relaxation if and only if
\begin{subequations}\label{eq:crit-less}
\begin{equation}
-\alpha_1(0)^2+\alpha_1(0)\alpha_2(0)-2\,\gamma_1(0)+\gamma_2(0) > 0 \qquad\mbox{ or }\qquad -\alpha_2(0)^2+\alpha_1(0)\alpha_2(0)-2\,\gamma_2(0)+\gamma_1(0) > 0\, ,
\label{eq:crit-less-2}
\end{equation}
and
\begin{equation}
-\alpha_2(0)^2+\alpha_1(0)\alpha_2(0)+\gamma_1(0)+\gamma_2(0) \leq 0\, ,
\label{eq:crit-less-3}
\end{equation}
and
\begin{equation}
\bigl[\alpha_1(0)-\alpha_2(0)\bigr]\,\bigl[\alpha_1(0)-2\alpha_2(0)\bigr] > 0\, .
\label{eq:crit-less-4}
\end{equation}
\end{subequations}
It undergoes a gradient catastrophe otherwise.
\end{prop}
\begin{proof}
Let $Q'$ denote the derivative of $Q$ with respect to $p$. We have
$$
Q'(p)=c_1+2\,c_2p+3\,p^2\,.
$$
As $\Delta_{Q}>0$, then $Q$ features three real roots, and the discriminant of $Q'$ is also positive. Therefore, the two roots of $Q'$ are also real. Write
$$
p_{\max}=\frac{-c_2-\sqrt{c_2^2-3c_1}}{3}
\qquad\mbox{and}\qquad
p_{\min}=\frac{-c_2+\sqrt{c_2^2-3c_1}}{3}
$$
for the roots of $Q'$. As
$$
Q''(p_{\max})=-2\,\sqrt{c_2^2-3c_1}<0
\qquad\mbox{and}\qquad
Q''(p_{\min})=2\,\sqrt{c_2^2-3c_1}>0\,,
$$
then $p_{\max}$ and $p_{\min}$ are the local maximum and minimum of $Q$, respectively. By Lemma \ref{lem:root-in-01}, the solution relaxes precisely when $p_{\min}$ lies strictly between $0$ and $1$.
Consider first when $p_{\min}$ is positive, that is, when $\sqrt{c_2^2-3\,c_1}$ exceeds $c_2$. This is verified when $c_2$ is negative, whatever the value of $c_1$ or when $c_2$ is non-negative and $c_1$ is negative. In other words, $p_{\min}$ is positive if $c_1$ is negative, or $c_2$ is negative.
Consider next when $p_{\min}$ is less than $1$, that is, when $\sqrt{c_2^2-3c_1}<3+c_2$. This can never happen if $c_2<-3$. If $3+c_2$ is non-negative, we can square both sides of the inequality, getting $c_2^2-3c_1<9+6c_2+c_2^2$, that is, $c_1>-3-2\,c_2$. So $p_{\min}$ is less than $1$ precisely when $c_2\geq-3$, and, in addition, $c_1>-3-2\,c_2$. Therefore, $p_{\min}$ lies in $(0,1)$, and the solution features asymptotic relaxation, if and only if all the above constraints are verified.
Substituting $c_1,c_2$ from (\ref{eq:cubic-coeffs}) and clearing the denominator $\gamma_2(0)$, turns the condition $c_1<0$ into $-\alpha_1(0)^2+\alpha_1(0)\alpha_2(0)-2\gamma_1(0)+\gamma_2(0)>0$, and $c_2<0$ into $\alpha_1(0)\alpha_2(0)-\alpha_2(0)^2+\gamma_1(0)-2\gamma_2(0)>0$. Together these give (\ref{eq:crit-less-2}). Likewise $c_2\geq-3$ becomes (\ref{eq:crit-less-3}), and $c_1>-3-2\,c_2$ becomes $-\alpha_1(0)^2+3\,\alpha_1(0)\alpha_2(0)-2\,\alpha_2(0)^2>0$, which factors as (\ref{eq:crit-less-4}).
\end{proof}
\begin{remark}
Under the hypothesis of Proposition \ref{prop:criterion-less}, condition (\ref{eq:crit-less-4}) is automatically satisfied.
\end{remark}
It is also immediate to verify the following proposition.
\begin{prop}
\label{prop:criterion-less-symmetric}
If $\gamma_1(0)=\gamma_2(0)$, conditions (\ref{eq:crit-less}) hold automatically whenever $\alpha_1(0)>0,\alpha_2(0)>0$. In other words, the criterion of Proposition \ref{prop:criterion-less} reduces to $\Delta_{Q}>0$.
\end{prop}
Finally, we can write a closed-form expression for the catastrophe time. From (\ref{eq:p-sol-t0}), we have $t_0=-\hat{f}(1)$, so, via (\ref{eq:p-sol-implicit}),
\begin{subequations}\label{eq:tcat-unequal}
\begin{equation}
t_c=\hat f(0)-\hat f(1)=\hat f(0)+t_0\, ,
\label{eq:tcat-unequal-a}
\end{equation}
where $\hat{f}(0)$ is obtained by setting $p=0$ in (\ref{eq:fhat}):
\begin{equation}
\hat{f}(0)=-\frac{\left[\alpha_1(0)-\alpha_2(0)\right]^3}{\gamma_2(0)^2}\sum_{\mathrm{cyc}}\left[\frac{2(c_0+p_1^3)}{(p_1-p_2)^3(p_1-p_3)^3}\log(-p_1)-\frac{p_1}{(p_1-p_2)^2(p_1-p_3)^2}\right]\, ,
\label{eq:fhat-zero}
\end{equation}
\end{subequations}
valid whenever the criterion above gives a catastrophe. As in Proposition \ref{prop:alpha-equal}, $t_c$ does not depend on $\mu_1(0)$ or $\mu_2(0)$. As $\Delta_{Q}$ approaches zero from within the catastrophe region, that is as $p_{\min}$ approaches zero from within the interval $(0,1)$, $t_c$ diverges as $t_c=O\bigl(|\Delta_{Q}|^{-3/2}\bigr)$. Conversely, as $\Delta_{Q}<0$ and either $\alpha_{1}(0)$ or $\alpha_{2}(0)$ approaches $-\infty$, the catastrophe time $t_{c}$ approaches zero.

\subsubsection{The case $\alpha_1(0)>\alpha_2(0)$}
\label{sec-alpha-greater}
Here $p$ increases monotonically from $1$ towards $\infty$, and the same reasoning as above applies with the roles of the two endpoints exchanged: a root $p_{k}$ of $Q$ lying beyond $1$ makes $\dot{p}$ vanish quadratically as $p$ approaches it, so $p(t)\to p_{k}$ only as $t\to\infty$ (relaxation), while the corresponding pole of $\hat{f}(p)$ at $p=p_{k}$ makes $t$ diverge, precisely as before, as $t=\hat{f}(p)+t_0$. On the contrary, if no root of $Q$ lies in $(1,\infty)$, then $Q(p)\neq0$ throughout, $\dot{p}$ never vanishes, and $p$ reaches infinity in finite time, as shown in Proposition \ref{prop:delta-sufficient} (gradient catastrophe). So relaxation occurs exactly when $Q$ has a real root in $(1,\infty)$.
\begin{lem}
\label{lem:root-in-1inf}
Let $\Delta_{Q}>0$, with $p_{\max}<p_{\min}$ as in Lemma \ref{lem:root-in-01}. Then $Q$ has a real root in $(1,\infty)$ if and only if $p_{\min}>1$.
\end{lem}
\begin{proof}
In analogy to what we did for Lemma \ref{lem:root-in-01}, we use $Q(1)>0$ and $Q(+\infty)=+\infty$ (same sign) in place of $Q(0)$ and $Q(1)$. If $p_{\min}\leq1$, then $Q$ is monotonic on $(1,\infty)$ past both critical points, with $Q(1)>0$ and $Q(p)>0$ for all $p>1$. Conversely, if $p_{\min}>1$, then $Q(1)>0>Q(p_{\min})$ forces a root in $(1,p_{\min})\subset(1,\infty)$ by the intermediate value theorem.
\end{proof}
\begin{prop}
\label{prop:criterion-greater}
Let $\gamma_1(0),\gamma_2(0)<0$, $\alpha_1(0)>\alpha_2(0)$, $\alpha_1(0)>0$, $\alpha_2(0)>0$, and $\Delta_{Q}>0$. The solution undergoes asymptotic relaxation if and only if
\begin{equation}
-\alpha_2(0)^2+\alpha_1(0)\alpha_2(0)+\gamma_1(0)+\gamma_2(0) > 0
\qquad\mbox{ or }\qquad
\left[\alpha_1(0)-2\alpha_2(0)\right]\,\left[\alpha_1(0)-\alpha_2(0)\right] < 0\, .
\label{eq:crit-greater}
\end{equation}
It undergoes a gradient catastrophe otherwise.
\end{prop}
\begin{proof}
By Lemma \ref{lem:root-in-1inf}, relaxation holds precisely when $p_{\min}$ exceeds $1$, that is, when $\sqrt{c_2^2-3\,c_1}>3+c_2$. If $c_2<-3$, then this holds automatically, since the left-hand side of the inequality can never be negative. If $c_2\geq-3$, both sides of the inequality are non-negative, so the inequality may be squared, giving $c_2^2-3\,c_1>9+6\,c_2+c_2^2$, that is, $c_1<-3-2\,c_2$. So $p_{\min}$ exceeds $1$ precisely when $c_2<-3$, or, failing that, when $c_1<-3-2\,c_2$.
Substituting $c_1$, $c_2$ from (\ref{eq:cubic-coeffs}) and clearing the denominator $\gamma_2(0)$, which is negative and so reverses each inequality, turns $c_2<-3$ into $\alpha_1(0)\alpha_2(0)-\alpha_2(0)^2+\gamma_1(0)+\gamma_2(0)>0$, and $c_1<-3-2\,c_2$ into $-\alpha_1(0)^2+3\,\alpha_1(0)\alpha_2(0)-2\,\alpha_2(0)^2<0$, which factors as the second term of (\ref{eq:crit-greater}).
\end{proof}
Similarly to what we have done for the case $\alpha_{1}(0)<\alpha_{2}(0)$, we show that the criterion (\ref{eq:crit-greater}) simplifies significantly when $\gamma_{1}(0)=\gamma_{2}(0)$.
\begin{prop}
\label{prop:criterion-greater-symmetric}
If $\gamma_1(0)=\gamma_2(0)$, condition (\ref{eq:crit-greater}) holds automatically whenever $\Delta_{Q}>0$, $\alpha_1(0)>0$, $\alpha_2(0)>0$. In other words, the criterion of Proposition \ref{prop:criterion-greater} reduces to $\Delta_{Q}>0$.
\end{prop}
\begin{proof}
The scaling $\alpha_i\mapsto\lambda\alpha_i$, $\gamma_i\mapsto\lambda^2\gamma_i$, $t\mapsto t/\lambda$ leaves (\ref{eq:sigma-system}) invariant for any $\lambda>0$, and $\Delta_{Q}$ scales homogeneously (degree $6$) under it, so its sign, and hence every statement below, is unaffected by fixing $\gamma_1(0)=\gamma_2(0)=-1$.
As before, we write $D=\alpha_1(0)-\alpha_2(0)>0$, $S=\alpha_1(0)+\alpha_2(0)$. Let $r$ be the ratio  $r=S/D$. As $\alpha_2(0)>0$, $S>D$, so $r>1$. In these variables, we have
$$
\Delta_{Q}=3\,g\left(D^2;r\right)
$$
where the function $g$ is defined as follows:
$$
g(z;r) = -512 + 16\,\left(r^2+12\right)\,z-8\,\left(5\,r^2+3\right)\,z^2+\left(r^2-1\right)^2\,z^3\,.
$$
The logical negation of the two terms of (\ref{eq:crit-greater}) translates into
$$
1<r\leq3
\qquad\mbox{and}\qquad
0<z\leq4/(r-1)\,.
$$
We show $g\left(D^2;r\right)<0$ throughout this region of $r$ and $z$, hence proving the statement of the proposition by logical negation.
The discriminant of $g$ with respect to $z$ is
$$
\Delta_{z}(g)=-16384\,r^4\,\left(r^2+27\right)^3\,,
$$
which vanishes (for real $r$) only at $r=0$. As the discriminant is never zero for $r\in(0,3]$, the polynomial $g(z;r)$ has no repeated root in $z$ anywhere in $(0,4/(r-1)]$, and the number of its real roots inside any fixed $z$-interval cannot change as $r$ varies continuously over $(1,3]$. Moreover $g(0;r)=-512$ identically, and
$$
g\left(\frac{4}{r-1};r\right)=64\,\frac{-27+27\,r-18\,r^2+2\,r^3}{(r-1)^2}\,,
$$
whose numerator is a cubic in $r$ with a single real root, at $r\approx7.43$, strictly outside $(1,3]$. Therefore, $g$ vanishes at neither endpoint of $\bigl(0,4/(r-1)\bigr)$ for any $r$ in this range either. Consequently, the number of roots in $z$ of $g(z;r)$ inside $\bigl(0,4/(r-1)\bigr)$ is the same for all $r\in(1,3]$, and it may be computed at a chosen representative value. We choose $r=2$, where we have that $g(z;2)=-512 + 256\,z - 184\,z^2 + 9\,z^3$ has exactly one real root, at $z\approx19.1$, outside $(0,4)$. Thus, zero roots lie inside the interval, for all $r\in(1,3]$. Since $g(0;z)=-512<0$ and $g$ has no root to cross in $\bigl(0,4/(r-1)\bigr)$, then $g(z;r)$ stays negative throughout, for all $r\in(1,3]$. This entails that $\Delta_{Q}<0$ whenever both terms of (\ref{eq:crit-greater}) fail, proving the proposition by logical negation of its statement.
\end{proof}
For the catastrophe time, the coefficients of $\log(p-p_i)$ in (\ref{eq:fhat}) are (up to the overall prefactor) the residues of $p^2/Q(p)^2$ at its three double poles, and their sum vanishes identically -- a direct computation, for arbitrary $p_1,p_2,p_3$, independent of any specific values. Writing $\log(p-p_i)=\log p+\log(1-p_i/p)$, the vanishing sum cancels the divergent $\log p$ term entirely, and each remaining $\log(1-p_i/p)\to\log1=0$ as $p\to\infty$. The rational terms in (\ref{eq:fhat}) vanish too. Hence
\begin{equation}
\lim_{p\to\infty}\hat f(p)=0\, ,
\end{equation}
exactly, and therefore
\begin{equation}
t_c=\lim_{p\to\infty}\hat f(p)-\hat f(1)=-\hat f(1)=t_0\, ,
\label{eq:tcat-greater}
\end{equation}
valid whenever the criterion above gives a catastrophe: the catastrophe time coincides exactly with $t_0$, with no further evaluation of $\hat f$ required. As before, $t_c$ does not depend on $\mu_1(0)$ or $\mu_2(0)$. As $\Delta_{Q}\to0$ from within the catastrophe region -- $p_{\min}$ approaching the boundary of $(1,\infty)$ -- $t_c\to\infty$, precisely as $t_c=O\bigl(|\Delta_{Q}|^{-3/2}\bigr)$.

\section{Identical initial supports}
\label{sec-identical-supports}
\begin{figure}[!ht]
\centering
\begin{subfigure}{0.495\textwidth}
\centering
\includegraphics[width=\textwidth]{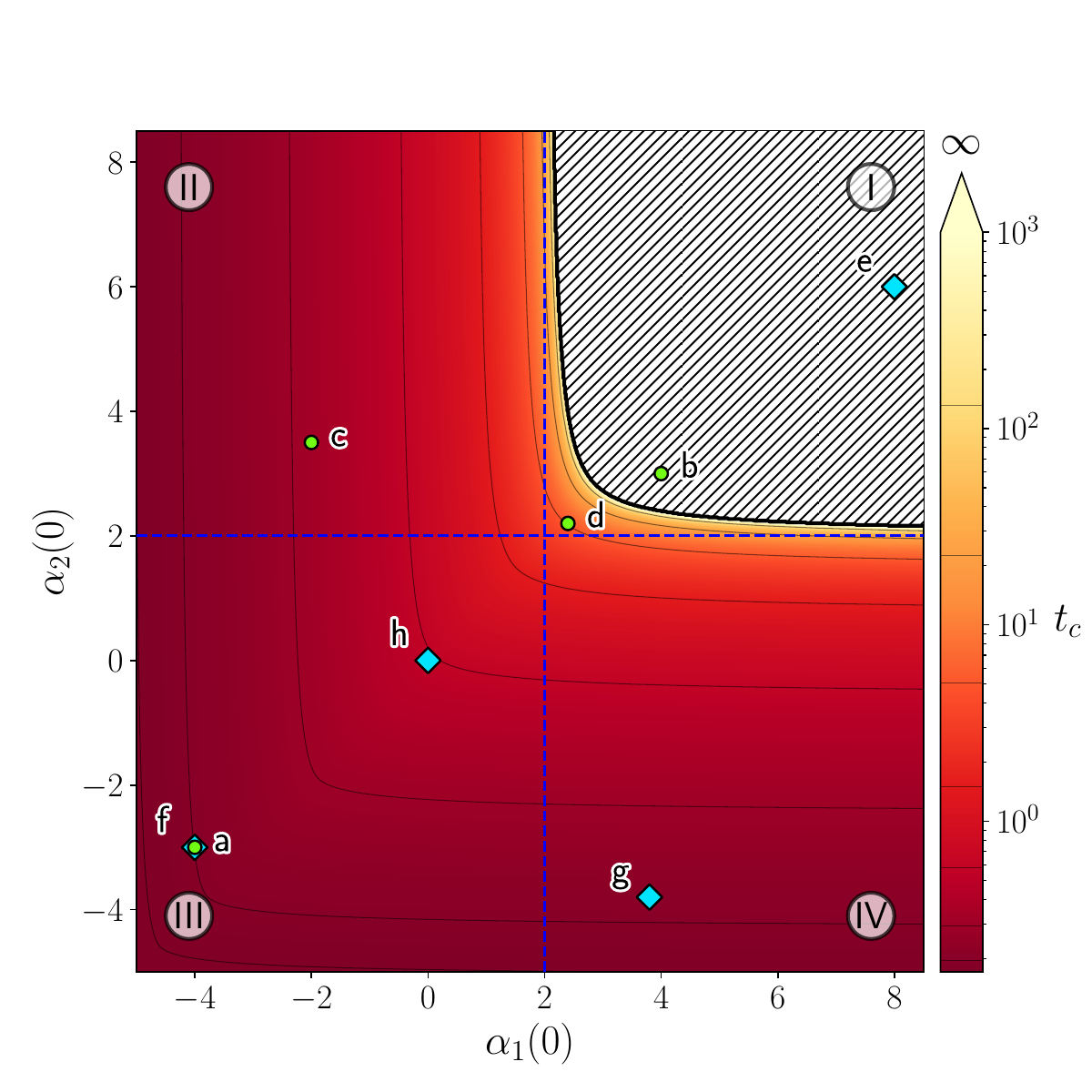}
\caption{}
\label{fig:region-plot-symmetric}
\end{subfigure}
\hfill
\begin{subfigure}{0.495\textwidth}
\centering
\includegraphics[width=\textwidth]{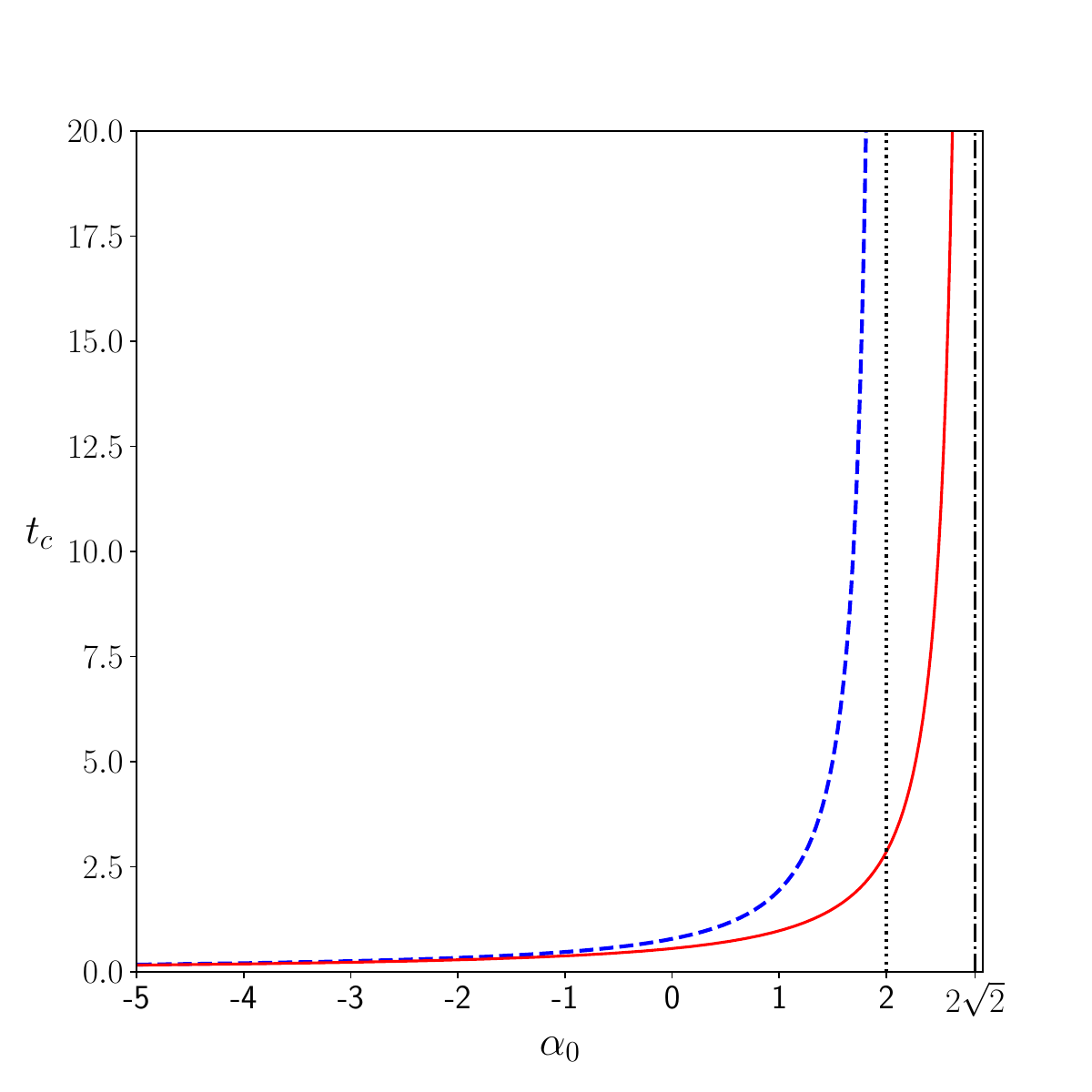}
\caption{}
\label{fig:tcat-alpha0}
\end{subfigure}
\caption{\textit{On the left} (Figure \ref{fig:region-plot-symmetric}): Time of catastrophe (colour gradient) as a function of the initial chirps versus asymptotic relaxation (striped region) in the $(\alpha_1(0),\alpha_2(0))$-plane when $\gamma_1(0)=\gamma_2(0)=-1$. The black threshold curve (corresponding to the infinite, continuous branch of $\Delta_{Q}=0$ in the first quadrant) divides the plane in two regions: below the curve, $\Delta_Q<0$ in the first quadrant and the solution features a catastrophe of the gradient; above the curve (diagonally striped region), $\Delta_Q>0$ and the solution features asymptotic relaxation (in general, for $\gamma_1(0)\neq\gamma_{2}(0)$ one would need to verify conditions (\ref{eq:crit-less}) or (\ref{eq:crit-greater})). The time of catastrophe approaches zero as either $\alpha_{1}(0)$ or $\alpha_{2}(0)$ approaches $-\infty$, whereas it diverges as the point $(\alpha_{1}(0),\alpha_{2}(0))$ approaches the threshold curve from within the catastrophe region. For our choice $\gamma_1(0)=\gamma_2(0)=-1$, the threshold curve features vertical and horizontal asymptotes at $\alpha_1(0)=2$ and $\alpha_2(0)=2$, respectively, and meets the diagonal at $\alpha_1(0)=\alpha_2(0)=2\sqrt2$. \colorblue{These asymptotes, drawn as dashed blue lines, divide the plane into four sectors, labelled I to IV, which classify the behaviour the two fields would display in the absence of coupling: both fields would relax in Sector I, both would undergo a gradient catastrophe in Sector III, while in Sector II the first would catastrophise and the second relax, and vice versa in Sector IV. Together with the threshold curve, they delimit the four domains (a)--(d) listed at the end of this Section.} Points a, b, c and d (lime green circles) mark the four regimes illustrated in Figure \ref{fig:regimes} (Figures \ref{fig:regime-a}, \ref{fig:regime-b}, \ref{fig:regime-c} and \ref{fig:regime-d}, respectively). Points e, f, g and h (cyan blue diamonds) mark the four cases investigated numerically and shown in Figure \ref{fig:dispersive1} (Figures \ref{fig:dispersive1-e}, \ref{fig:dispersive1-f}, \ref{fig:dispersive2-g} and \ref{fig:dispersive2-h}, respectively). \textit{On the right} (Figure \ref{fig:tcat-alpha0}): The catastrophe time $t_c$ as a function of $\alpha_0$ (red continuous line) for $\alpha_1(0)=\alpha_2(0)=\alpha_0$, $\gamma_1(0)=\gamma_2(0)=-1$, diverging for $\alpha_{0} = 2\sqrt{2}$, formula (\ref{eq:tcat-equal}), overlapped with the catastrophe time as a function of the chirp for a parabolic initial condition of density $\rho=1-x^2$ and phase $u=\alpha_{0}\,x$ for the scalar focussing NLS equation studied in \cite{DORS23}, diverging at $\alpha_{0}=2$ (blue dashed line). The red curve is always below the blue one, in line with our general observation that the catastrophe for the coupled system (vector NLS) occurs earlier than in the absence of the coupling (scalar NLS).}
\end{figure}

As mentioned in the Introduction and following Remark \ref{rem:validity}, and aiming at comparing the dispersionless Talanov dynamics against the fully dispersive case, in this section we focus on initial data with identical initial supports. In analogy with the semiclassical comparison carried out in \cite{DORS23} for the scalar problem, the dispersionless solution is expected to provide insight into the evolution of a neighbourhood of the parabolic maxima in the fully dispersive dynamics.
Throughout this section, by dispersionless dynamics we mean the following two-step construction: we solve the system of ODEs (\ref{eq:ODEs-alpha-gamma-mu}) for $\alpha_i(t),\gamma_i(t),\mu_i(t)$ -- quantities defined for all $t$, independently of $x$ -- and use them to form the two quadratic profiles $\gamma_i(t)x^2+\mu_i(t)$, defined for all $x$. We then restrict each one to where it is non-negative, as in (\ref{eq:compact-rho}), discarding it elsewhere. As discussed following Remark \ref{rem:validity}, this construction solves the coupled system (\ref{eq-dMan}) only where both restricted densities are simultaneously non-zero, and so can be expected to approximate the true dispersionless dynamics only in a neighbourhood of the parabolic maxima at $x=0$, where the two supports overlap. It becomes an exact solution of the dispersionless system for compactly-supported, parabolic initial data only when the initial chirps and the initial supports are identical (Remark \ref{rem:exact-identical}).
The support of the initial density $\rho_i(x,0)=\gamma_i(0)x^2+\mu_i(0)$, with $\gamma_i(0)<0$ and $\mu_i(0)>0$, $i=1,2$, is the interval
\begin{equation}\label{eq:initial-interval}
\left[-L_i(0),L_i(0)\right]
\quad\mbox{with}\quad
L_i(0)=\sqrt{-\frac{\mu_i(0)}{\gamma_i(0)}}\,,
\qquad i=1,2\,,
\end{equation}
and the two supports coincide, at $t=0$, exactly when $\mu_1(0)/\gamma_1(0)=\mu_2(0)/\gamma_2(0)$, in other words, whenever the two initial $\mu$'s are proportional to the two initial $\gamma$'s through the same constant of proportionality.
This condition constrains only the $\mu$'s relative to the $\gamma$'s, and not the $\alpha$'s or $\gamma$'s themselves. Indeed, (\ref{eq:sigma-system}) shows that $\mu_1$ and $\mu_2$ are entirely decoupled from $\alpha_1$, $\alpha_2$, $\gamma_1$, $\gamma_2$. Consequently, identical initial supports place no restriction at all on $\alpha_1(0)$, $\alpha_2(0)$, $\gamma_1(0)$, $\gamma_2(0)$: given any choice of these four quantities, setting $\mu_1(0)$, $\mu_2(0)$ proportional to $\gamma_1(0)$, $\gamma_2(0)$, respectively, through a common constant gives identical initial supports. Therefore, the entire space of dynamics described in Sections \ref{sec-alpha-equal}--\ref{sec-alpha-unequal} is relevant for the case of identical initial supports.
Within this family, we focus throughout on the convenient choice $\gamma_1(0)=\gamma_2(0)=-1,\mu_1(0)=\mu_2(0)=1$: numerical exploration confirms it is representative of the entire phenomenology of the identical-initial-supports case.
For identical chirps, $\alpha_1(0)=\alpha_2(0)=\alpha_0$, Proposition \ref{prop:alpha-equal} specialises directly, with $\gamma_0=\gamma_1(0)+\gamma_2(0)=-2$, and $A=\alpha_0^2-8$. We have
\begin{subequations}\label{eq:solution-equal-symmetric}
\begin{equation}\label{eq:alphagammamu-equal-symmetric}
\alpha_1=\alpha_2=\frac{\dot{\sigma}}{\sigma}\,,\quad
\gamma_1=\gamma_2=-\frac{1}{\sigma^3}\,,\quad
\mu_1=\mu_2=\frac{1}{\sigma}\,,
\end{equation}
where $\sigma\equiv\sigma(t)$ satisfies $\sigma(0)=1$, and, for all $t>0$, is given implicitly as the solution of the following equation:
\begin{equation}
f(\sigma)^2=A^3(t-t_0)^2\, ,
\end{equation}
with
\begin{equation}
f(\sigma)=\sigma\sqrt{A}\,\sqrt{A+\frac{8}{\sigma}}-8\operatorname{acoth}\!\left(\frac{\sqrt A}{\sqrt{A+\frac{8}{\sigma}}}\right)\, ,
\quad\mbox{and}\quad
t_0=-\frac{\alpha_0}{A}+\frac{8}{A^{3/2}}\operatorname{atanh}\!\left(\frac{\alpha_0}{\sqrt A}\right)\,.
\label{eq:sigma-equal-symmetric}
\end{equation}
\end{subequations}
For different chirps, $\alpha_1(0)\neq\alpha_2(0)$, after setting $\gamma_1(0)=\gamma_2(0)=-1$ in (\ref{eq:cubic-coeffs}), the coefficients of the cubic polynomial $Q(p)=c_0+c_1p+c_2p^2+p^3$ in (\ref{eq:cubic}) become
\begin{subequations}\label{eq:solution-general-symmetric}
\begin{equation}\label{eq:cubic-coeffs-symmetric}
c_0 = 1\,,
\qquad
c_1 = \alpha_1(0)^2 - \alpha_1(0)\,\alpha_2(0) - 1\,,
\qquad
c_2 = \alpha_2(0)^2 - \alpha_1(0)\,\alpha_2(0) - 1\,.
\end{equation}
Then, Proposition \ref{prop:alpha-unequal} yields:
\begin{align}
\alpha_1&=\frac{\left(2+c_1p-p^3\right)\,Q(p)}{\left[\alpha_1(0)-\alpha_2(0)\right]^3\,p^3}\,,&
\alpha_2&=\frac{\left(1-c_2p^2-2p^3\right)\,Q(p)}{\left[\alpha_1(0)-\alpha_2(0)\right]^3\,p^3}\, ,\\
\gamma_1&=-\frac{|Q(p)|^3}{\left[\alpha_1(0)-\alpha_2(0)\right]^6\,|p|^3\,p^3}\,,&
\gamma_2&=-\frac{|Q(p)|^3}{\left[\alpha_1(0)-\alpha_2(0)\right]^6\,|p|^3}\,,\\
\mu_1&=\frac{|Q(p)|}{\left[\alpha_1(0)-\alpha_2(0)\right]^2\,|p|\,p}\,,&
\mu_2&=\frac{|Q(p)|}{\left[\alpha_1(0)-\alpha_2(0)\right]^2\,|p|}\,,
\label{eq:reconstruction-symmetric}
\end{align}
where $p\equiv p(t)$ satisfies $p(0)=1$ and, for all $t>0$, is given implicitly as the solution of the following equation:
\begin{equation}
\hat{f}(p) = t-t_{0}\,,
\end{equation}
with
\begin{equation}
\hat{f}(p)=-\left[\alpha_1(0)-\alpha_2(0)\right]^3\,
\sum_{\mathrm{cyc}}\left[\frac{2\,(1+p_1^3)}{(p_1-p_2)^3\,(p_1-p_3)^3}\log(p-p_1)+\frac{p_1^2}{(p_1-p_2)^2\,(p_1-p_3)^2\,(p-p_1)}\right]\,,
\label{eq:fhat-symmetric}
\end{equation}
and $t_0=-\hat{f}(1)$, while $p_1$, $p_2$ and $p_3$ are the three roots of $Q$.
\end{subequations}
When $\gamma_1(0)=\gamma_2(0)=-1$ and $\mu_1(0)=\mu_2(0)$, the conditions in Propositions \ref{prop:criterion-less-symmetric} and \ref{prop:criterion-greater-symmetric} simplify: the solution can be shown to undergo asymptotic relaxation if and only if $\Delta_{Q}>0$ and $\alpha_1(0),\alpha_2(0)>0$, while featuring a blow-up at finite time otherwise. Therefore, the first-quadrant, infinite, continuous branch of the curve in the $(\alpha_{1}(0),\alpha_{2}(0))$-plane implicitly defined, via (\ref{eq:discriminant-DeltaQ}), as $\Delta_{Q}=0$, with
\begin{equation}\label{eq:discriminant-DeltaQ-symmetric}
\Delta_{Q}=3\left[D^6-(2S^2+24)D^4+(S^4-40S^2+192)D^2+16S^2-512\right]\,,
\end{equation}
where, as usual, $D=\alpha_1(0)-\alpha_2(0)$ and $S=\alpha_1(0)+\alpha_2(0)$, represents the threshold dividing the catastrophic behaviour from the relaxation behaviour: for a choice of $(\alpha_{1}(0),\alpha_{2}(0))$ below the curve, the solution undergoes gradient catastrophe while for a choice of $(\alpha_{1}(0),\alpha_{2}(0))$ above the curve, the solution undergoes asymptotic relaxation. On the curve, the solution catastrophises at infinity. This is illustrated in Figure \ref{fig:region-plot-symmetric}. In particular, when $\alpha_{1}(0)=\alpha_{2}(0)=\alpha_{0}$, the solution relaxes asymptotically whenever $\alpha_0\geq2\sqrt2$, as shown in Figure \ref{fig:tcat-alpha0}.
\begin{remark}
\label{rem:threshold-parametric}
By identifying $z=D=\alpha_{1}(0)-\alpha_{2}(0)$ in the expression (\ref{eq:discriminant-DeltaQ-symmetric}) for $\Delta_{Q}$, solving $\Delta_{Q}=0$ as a quadratic polynomial in $S$ and analysing the different branches of the solution, it is straightforward to show that the threshold curve $\Delta_{Q}=0$ admits an explicit parametrisation in the first quadrant of the $\left(\alpha_{1}(0),\alpha_{2}(0)\right)$-plane of the form:
\begin{subequations}\label{eq:threshold-param}
\begin{equation}
\alpha_1(0)=\frac{S(z)+z}{2}\,,\qquad \alpha_2(0)=\frac{S(z)-z}{2}\, ,
\label{eq:threshold-param-alphas}
\end{equation}
with
\begin{equation}
S(z)=\sqrt{\frac{z^4+20z^2-8+8(z^2+1)^{3/2}}{z^2}}\,,\qquad\mbox{for all}\quad z\in\mathbb{R}\,,
\label{eq:threshold-param-S}
\end{equation}
\end{subequations}
where $z<0$ corresponds to the part of the branch where $\alpha_{1}(0)<\alpha_{2}(0)$, whereas $z>0$ corresponds to the part of the branch where $\alpha_{1}(0)>\alpha_{2}(0)$. As illustrated in Figure \ref{fig:region-plot-symmetric}, the threshold curve $\Delta_{Q}=0$ is asymptotic to the lines $\alpha_1(0)=2$ and $\alpha_2(0)=2$: this can be verified by taking the limit of (\ref{eq:threshold-param}) as $z\to\pm\infty$, corresponding to the threshold values telling the relaxation from the catastrophe behaviour for the scalar case when $\gamma_{1}(0)=\gamma_{2}(0)=-1$, see \cite{DORS23}.
\end{remark}
The existence of asymptotes is a general phenomenon: for fixed generic $\gamma_{1}(0),\gamma_{2}(0)<0$, the infinite, continuous branch of $\Delta_{Q}=0$ in the first quadrant of the $\left(\alpha_{1}(0),\alpha_{2}(0)\right)$-plane always features a vertical asymptote at $\alpha_{1}(0)=2\,\sqrt{-\gamma_{1}(0)}$, as well as a horizontal asymptote at $\alpha_{2}(0)=2\,\sqrt{-\gamma_{2}(0)}$. These two latter lines correspond to the thresholds distinguishing the relaxation behaviour from the catastrophe behaviour for the scalar case. This can be understood by looking at the system of ODEs satisfied by the $\alpha$'s, $\gamma$'s and $\mu$'s, (\ref{eq:ODEs-alpha-gamma-mu}): if, say, $\alpha_{1}(0)$ becomes very large, then (\ref{eq:ODEs-gamma}) and (\ref{eq:ODEs-mu}) show that $\gamma_{1}$ and $\mu_{1}$ decay exponentially. Since the coupling between the two fields is only due to the presence of $\gamma_{1}$ and $\gamma_{2}$ in (\ref{eq:ODEs-alpha}), if $\gamma_{1}$ vanishes exponentially, then the two fields decouple, and each one of them behaves as the scalar system described in \cite{DORS23}.
As for the catastrophe time, $t_{c}$, we use equations (\ref{eq:tcat-equal}), (\ref{eq:tcat-unequal}) and (\ref{eq:tcat-greater}), depending on whether $\alpha_{1}(0)=\alpha_{2}(0)$, or $\alpha_{1}(0)<\alpha_{2}(0)$, or $\alpha_{1}(0)>\alpha_{2}(0)$, respectively:
\begin{equation}\label{eq:cat-time-equalgammas}
t_c=
\left\{
\begin{array}{ll}
\frac{f(0)-f(1)}{A^{3/2}}&
\mbox{for}\quad
\alpha_{1}(0)=\alpha_{2}(0)=\alpha_{0}
\quad\mbox{and}\quad
\alpha_{0}\in(-\infty,2\sqrt{2})\,,\alpha_{0}\neq-2\,\sqrt{2}\\
&\\
\frac{1}{3\sqrt{2}}&
\mbox{for}\quad
\alpha_{1}(0)=\alpha_{2}(0)=-2\,\sqrt{2}\\
&\\
\hat f(0)-\hat f(1)&
\mbox{for}\quad
\alpha_{1}(0)<\alpha_{2}(0)
\quad\mbox{and}\quad
\mbox{either}\;\;\Delta_{Q}<0\;\;\mbox{or}\;\;\alpha_{1}(0)\leqslant0\\
&\\
-\hat f(1)&
\mbox{for}\quad
\alpha_{1}(0)>\alpha_{2}(0)
\quad\mbox{and}\quad
\mbox{either}\;\;\Delta_{Q}<0\;\;\mbox{or}\;\;\alpha_{2}(0)\leqslant0
\end{array}
\right.
\end{equation}
where $A=\alpha_0^2-8$ as before in this Section, $f$ as in (\ref{eq:sigma-equal-symmetric}), $\hat{f}$ as in (\ref{eq:fhat-symmetric}), and $\Delta_{Q}$ as in (\ref{eq:discriminant-DeltaQ-symmetric}). A particularly neat value is obtained for $\alpha_{1}(0)=\alpha_{2}(0)=0$ yielding $t_{c}=\pi/(4\sqrt2)$. The behaviour of $t_{c}$ as a function of $\alpha_{0}\in(-\infty,2\sqrt{2})$ for the case $\alpha_{1}(0)=\alpha_{2}(0)=\alpha_{0}$ is illustrated in Figure \ref{fig:tcat-alpha0}, where it is compared against the behaviour of the catastrophe time for an identical initial profile in the scalar case studied in \cite{DORS23}, i.e. as a function of the chirp for a single parabolic initial condition of density $\rho=1-x^2$ and phase $u=\alpha_{0}\,x$ for the scalar focussing NLS equation, with $\alpha_{0}\in(-\infty,2)$.
\begin{figure}[!ht]
\centering
\begin{subfigure}{0.48\textwidth}
\centering
\includegraphics[width=\textwidth]{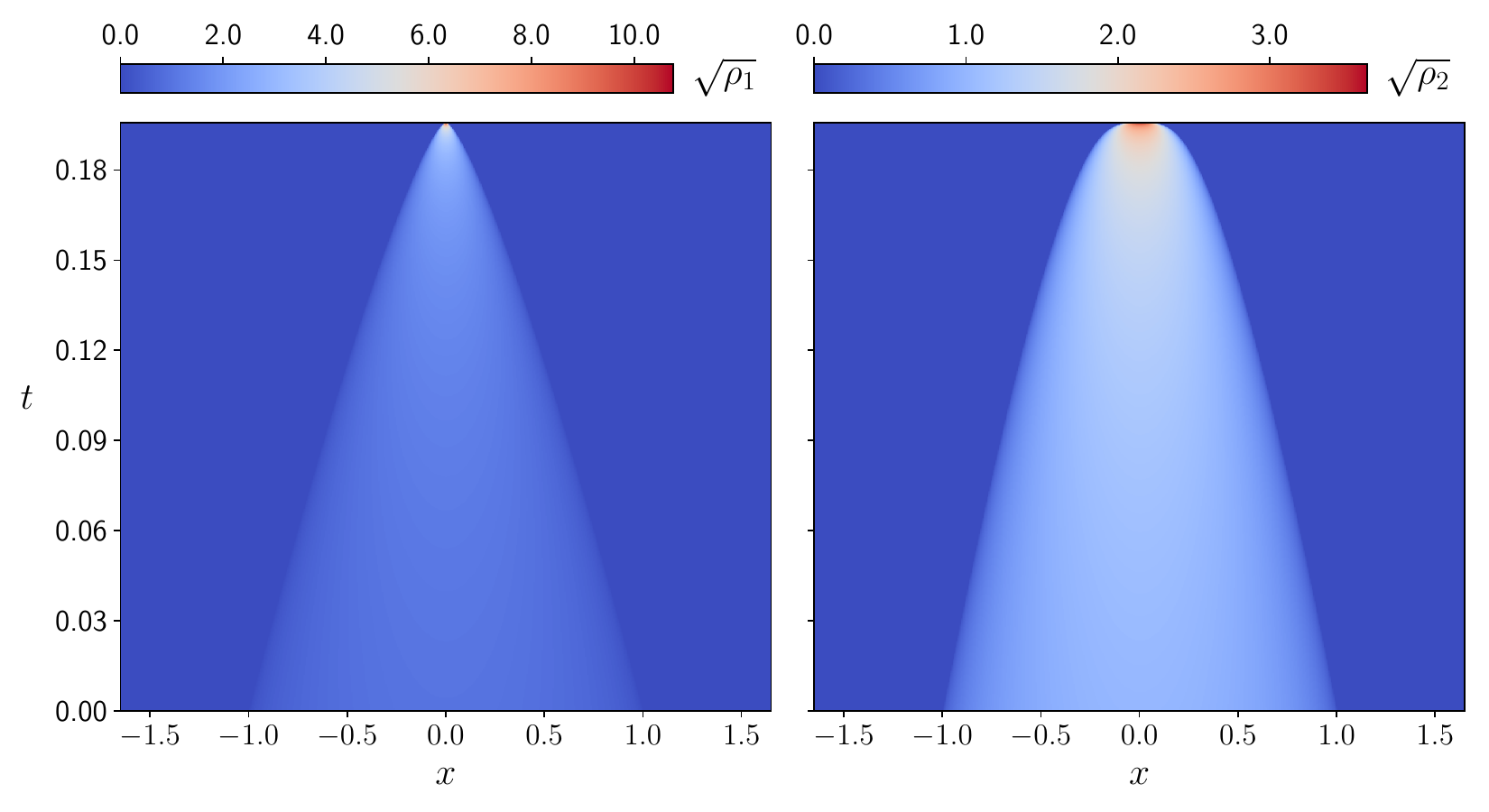}
\caption{$\alpha_1(0)=-4,\alpha_2(0)=-3$}
\label{fig:regime-a}
\end{subfigure}
\hfill
\begin{subfigure}{0.48\textwidth}
\centering
\includegraphics[width=\textwidth]{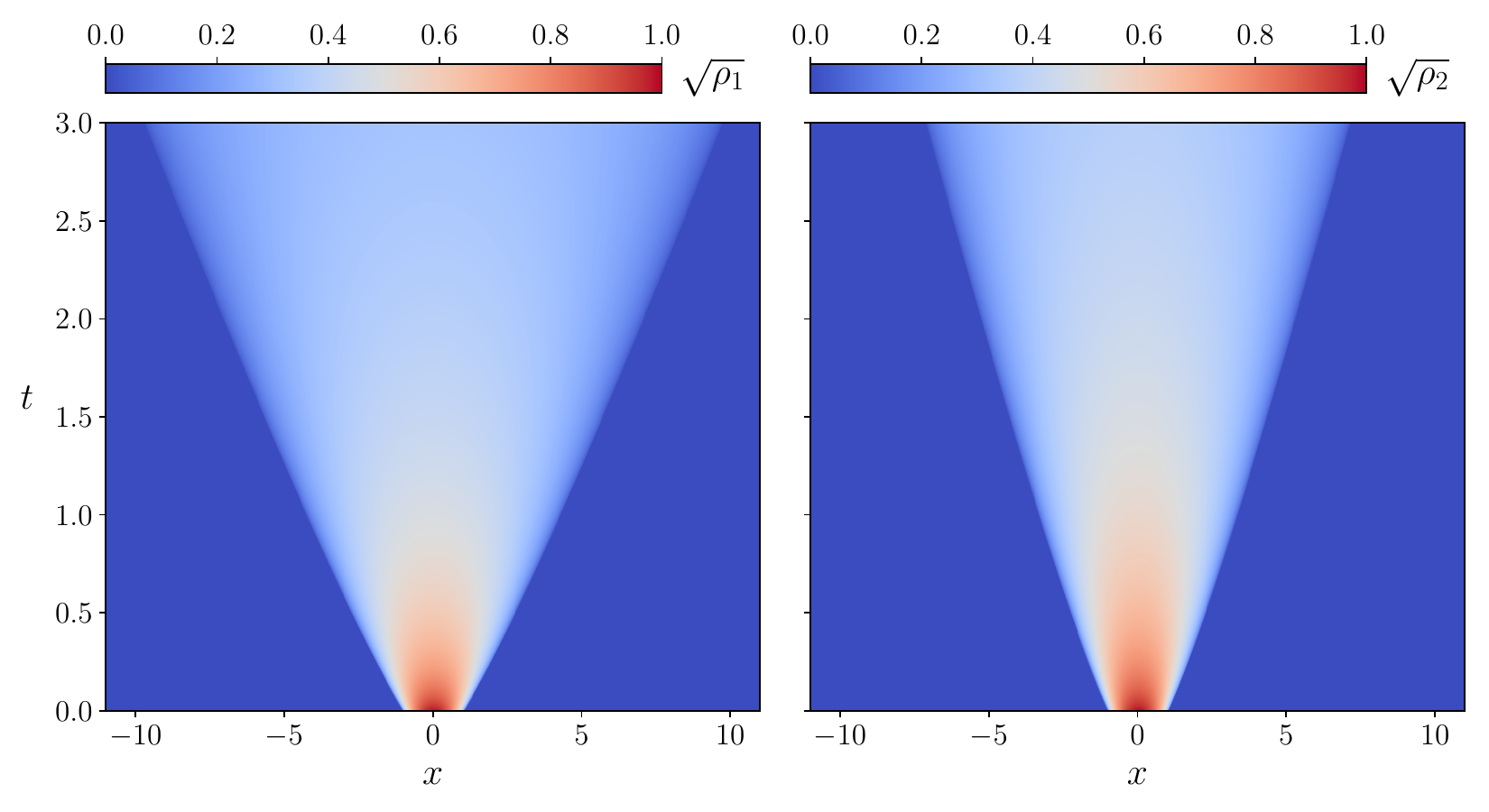}
\caption{$\alpha_1(0)=4,\alpha_2(0)=3$}
\label{fig:regime-b}
\end{subfigure}
\vspace{0.5cm}
\begin{subfigure}{0.48\textwidth}
\centering
\includegraphics[width=\textwidth]{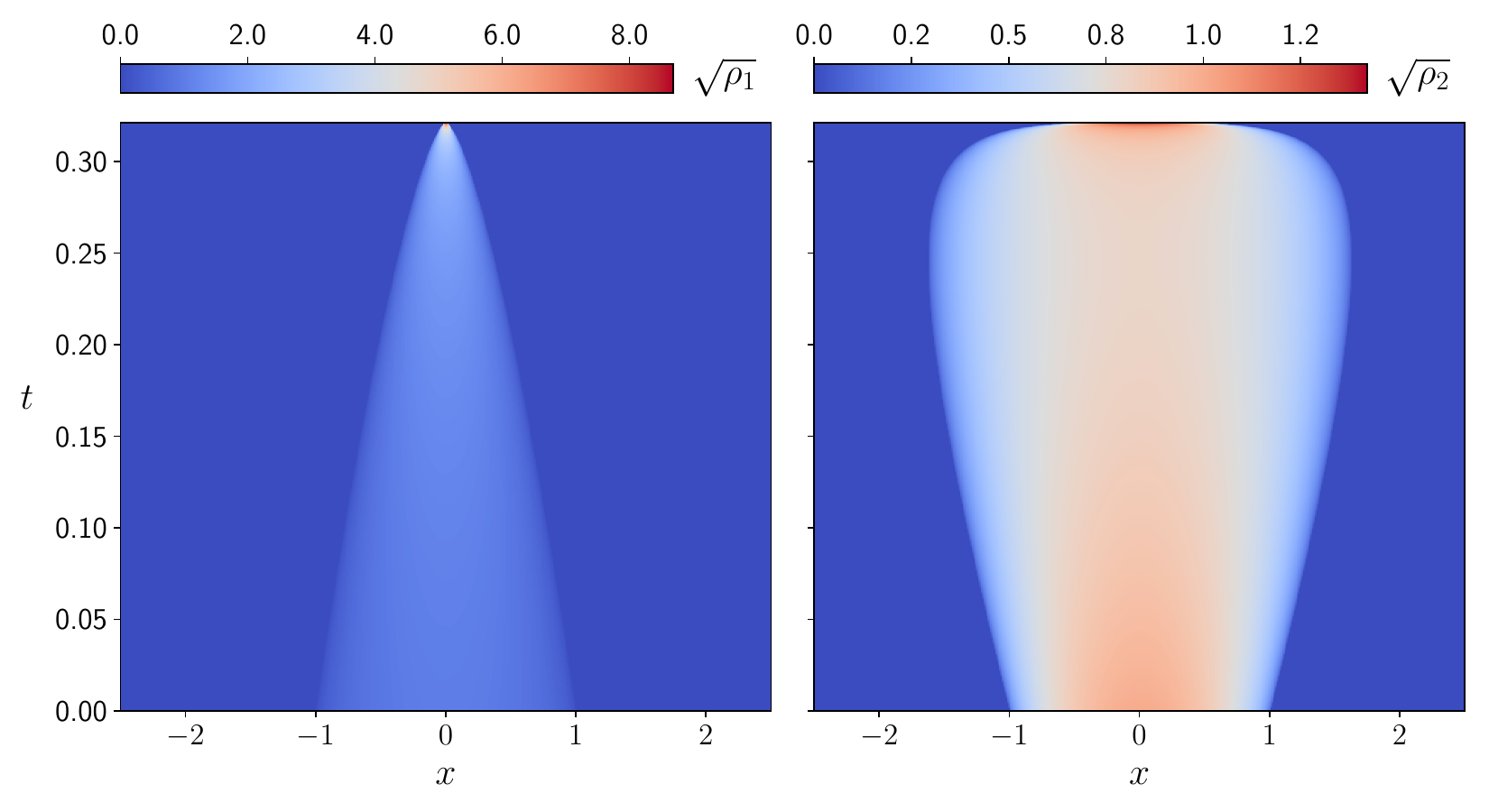}
\caption{$\alpha_1(0)=-2,\alpha_2(0)=3.5$}
\label{fig:regime-c}
\end{subfigure}
\hfill
\begin{subfigure}{0.48\textwidth}
\centering
\includegraphics[width=\textwidth]{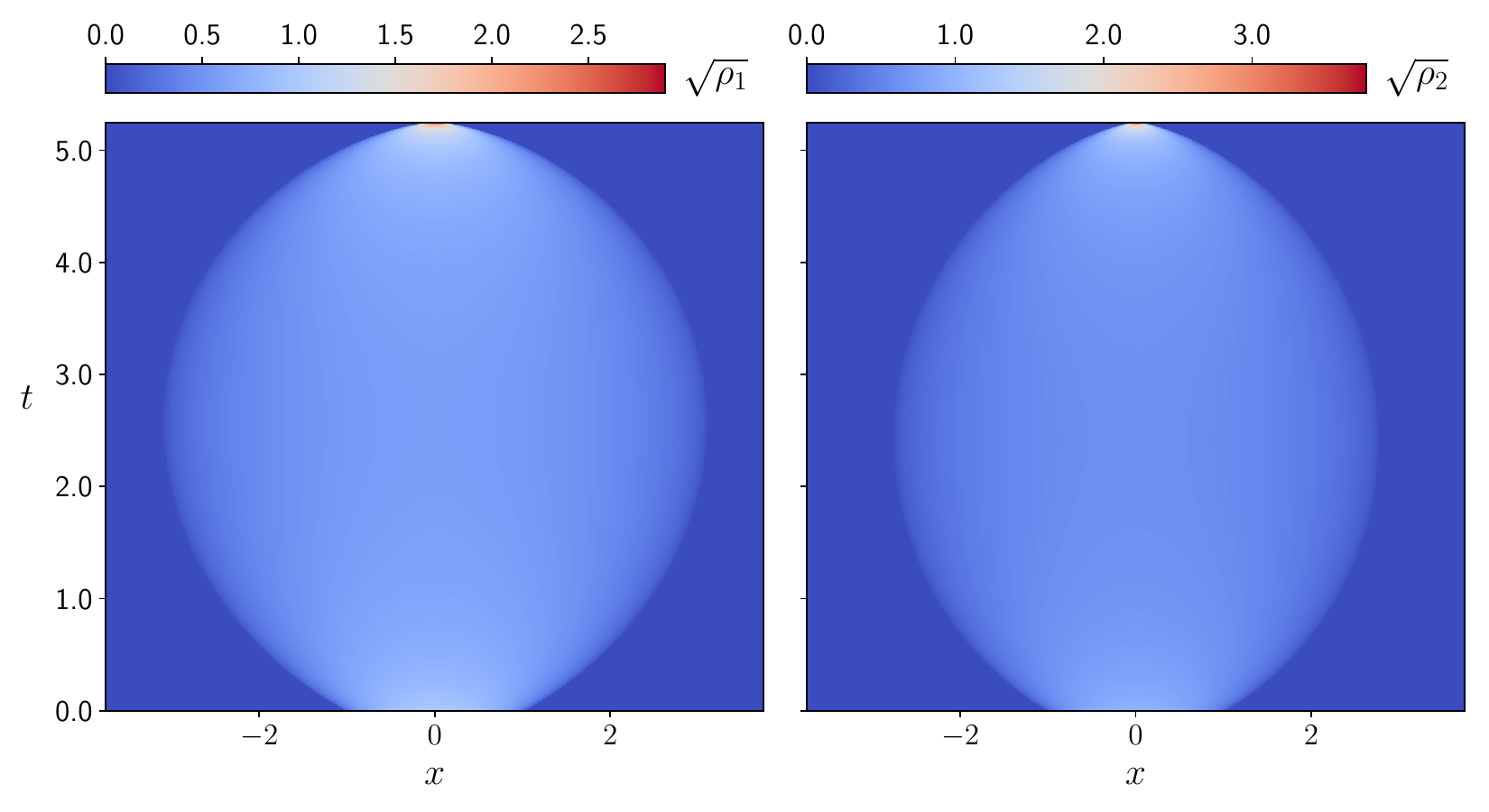}
\caption{$\alpha_1(0)=2.4,\alpha_2(0)=2.2$}
\label{fig:regime-d}
\end{subfigure}
\caption{Densities $\sqrt{\rho_1}$ and $\sqrt{\rho_2}$ (left and right panel of each pair, respectively), with $\gamma_1(0)=\gamma_2(0)=-1$ and $\mu_1(0)=\mu_2(0)=1$, for four choices of $\alpha_1(0)$ and $\alpha_2(0)$ illustrating the phenomenological range of the Talanov dynamics. Figures \ref{fig:regime-a}, \ref{fig:regime-b}, \ref{fig:regime-c} and \ref{fig:regime-d} correspond to points a, b, c and d, respectively, in Figure \ref{fig:region-plot-symmetric}. (a) Gradient catastrophe in both channels, agreeing with the decoupled prediction. (b) Relaxation in both channels, agreeing with the decoupled prediction. (c) Gradient catastrophe in both channels, whereas decoupled, the second channel would relax on its own (coupling drags the second channel into catastrophe). (d) Gradient catastrophe in both channels, even though both would relax if decoupled. In each of the four panels, only the common central region where both supports still overlap is expected to be representative of the true dispersionless dynamics (see the discussion after Remark \ref{rem:validity}).}
\label{fig:regimes}
\end{figure}
Comparing with the blow-up criterion identified in \cite{DORS23}, for fixed $\gamma_{1}(0)$ and $\gamma_{2}(0)$ the dispersionless solution to (\ref{eq-dMan}) displays four different behaviours, depending on the values of the chirps $\alpha_{1}(0)$ and $\alpha_{2}(0)$:
\begin{enumerate}[label=(\alph*)]
\item if $\alpha_{1}(0)<2\,\sqrt{-\gamma_{1}(0)}$ and $\alpha_{2}(0)<2\,\sqrt{-\gamma_{2}(0)}$, then, the parabolic initial conditions in the two fields of the dispersionless vector NLS system evolve into a simultaneous catastrophe at finite time, and they would catastrophise (possibly at different times) even if the two fields were decoupled (dispersionless scalar NLS). We observe numerically that the catastrophe time for the coupled system is shorter than the minimum of the catastrophe time for each individual evolution if the two fields were decoupled. This case is illustrated in Figure \ref{fig:regime-a} for $\gamma_{1}(0)=\gamma_{2}(0)=-1$ (corresponding to point a in Figure \ref{fig:region-plot-symmetric});
\item if $\alpha_{1}(0)>2\,\sqrt{-\gamma_{1}(0)}$, $\alpha_{2}(0)>2\,\sqrt{-\gamma_{2}(0)}$, $\Delta_{Q}>0$ and (\ref{eq:crit-less}) or (\ref{eq:crit-greater}) are satisfied, then, the parabolic initial conditions in the two fields of the dispersionless vector NLS system evolve into asymptotic relaxation, and they would relax asymptotically even if the two fields were decoupled (dispersionless scalar NLS). This case is illustrated in Figure \ref{fig:regime-b} for $\gamma_{1}(0)=\gamma_{2}(0)=-1$ (corresponding to point b in Figure \ref{fig:region-plot-symmetric}): for this choice of $\gamma_1(0)$,$\gamma_2(0)$, (\ref{eq:crit-less}) or (\ref{eq:crit-greater}) are automatically satisfied;
\item if $\alpha_{1}(0)<2\,\sqrt{-\gamma_{1}(0)}$ and $\alpha_{2}(0)>2\,\sqrt{-\gamma_{2}(0)}$, or, respectively, if $\alpha_{1}(0)>2\,\sqrt{-\gamma_{1}(0)}$ and $\alpha_{2}(0)<2\,\sqrt{-\gamma_{2}(0)}$, then, the parabolic initial conditions in the two fields of the dispersionless vector NLS system evolve into a simultaneous catastrophe at finite time. However, in this case, if the two fields were decoupled (dispersionless scalar NLS), the first parabolic initial condition would have evolved into a gradient catastrophe at finite time (resp. would have relaxed asymptotically), while the second one would have relaxed asymptotically (resp. would have evolved into a gradient catastrophe at finite time). In other words, in this scenario, the field doomed to catastrophise at finite time (if the two fields were decoupled) forces a (simultaneous) catastrophe in the second field as well, even if the latter were supposed to relax asymptotically, had the fields been decoupled. This case is illustrated in Figure \ref{fig:regime-c} for $\gamma_{1}(0)=\gamma_{2}(0)=-1$ (corresponding to point c in Figure \ref{fig:region-plot-symmetric});
\item if $\alpha_{1}(0)>2\,\sqrt{-\gamma_{1}(0)}$, $\alpha_{2}(0)>2\,\sqrt{-\gamma_{2}(0)}$ and $\Delta_{Q}<0$, then, the parabolic initial conditions in the two fields of the dispersionless vector NLS system evolve into a simultaneous catastrophe at finite time. However, in this case, the two parabolic initial conditions would evolve into asymptotic relaxation, had the fields been decoupled (dispersionless scalar NLS). This case shows that the coupling alone between $\rho_{1}$ and $\rho_{2}$ is sufficient to induce a catastrophe even when the two separate fields would undergo asymptotic relaxation. This case is illustrated in Figure \ref{fig:regime-d} for $\gamma_{1}(0)=\gamma_{2}(0)=-1$ (corresponding to point d in Figure \ref{fig:region-plot-symmetric}).
\end{enumerate}

\colorblue{It is also natural to inquire whether the four behaviours described above can be characterised in terms of the conserved quantities (\ref{eq:conserved-quantities}). Let $L(0)=L_{1}(0)=L_{2}(0)$ denote the half-width of the common initial support, so that $\mu_{i}(0)=-\gamma_{i}(0)L(0)^2$, and recall $\gamma_{0}=\gamma_{1}(0)+\gamma_{2}(0)$ from Proposition \ref{prop:alpha-equal}. The masses are $M_{i}=-\tfrac{4}{3}\gamma_{i}(0)L(0)^3$, each separately conserved, so that $\gamma_{0}$ is the total mass of the two fields up to a positive factor. The momentum $P$ vanishes by parity. The energy of the initial datum is
\begin{equation*}
H=-\frac{2}{15}\left[\alpha_{1}(0)^2\gamma_{1}(0)+\alpha_{2}(0)^2\gamma_{2}(0)+4\gamma_{0}^2\right]L(0)^5\, ,
\end{equation*}
whereas the energy of the $i$-th field in the absence of coupling is $H_{i}=-\tfrac{2}{15}\gamma_{i}(0)\left(\alpha_{i}(0)^2+4\gamma_{i}(0)\right)L(0)^5$. The latter is negative whenever $|\alpha_{i}(0)|<2\sqrt{-\gamma_{i}(0)}$. These quantities are related by $H-(H_{1}+H_{2})=-\tfrac{16}{15}\gamma_{1}(0)\gamma_{2}(0)L(0)^5$, which is strictly negative: the coupling lowers the energy by an amount bilinear in $\gamma_{1}(0)$ and $\gamma_{2}(0)$, determined by neither field in isolation. In particular, $H$ can be negative while $H_{1}$ and $H_{2}$ are both positive. Therefore, these relations alone cannot determine which of the four behaviours occurs, whereas this is governed by the criterion established above.}

\section{Dispersive effects on parabolic initial data and rogue wave formation}
\label{sec-dispersive}

The goal of the present section is to qualitatively compare the numerical solution of the full dispersive vector NLS system (\ref{eq-Man-NLS}) with the semianalytical solution of its dispersionless counterpart (\ref{eq-dMan}) constructed in the previous sections\colorblue{. In particular, we will show that,} in the semiclassical limit corresponding to weak dispersion, the dispersionless dynamics drives the evolution of the local parabolic maxima in the case of identical supports in the two coupled fields. Here we consider the initial condition (\ref{eq:compact-rho}) for both the dispersive and the dispersionless model, with $\mu_1(0)=\mu_2(0)=1$ and $\gamma_1(0)=\gamma_2(0)=-1$, that is,
\begin{equation}
\psi_k(x,0)=\psi_k^{(0)}(x)=\sqrt{1-x^2}\,\exp\left(\frac{i\alpha_k(0)}{2\epsilon}x^2\right)\, , \qquad k=1,2\, ,
\label{eq:disp-initial-condition}
\end{equation}
for different choices of the two chirps $\alpha_1(0),\alpha_2(0)$ and for small dispersion $\epsilon>0$. In this case, the dispersionless solution will undergo asymptotic relaxation or a gradient catastrophe at a finite time $t_c$, according to the criterion established in Section \ref{sec-identical-supports}, with $t_c$ given explicitly by (\ref{eq:cat-time-equalgammas}).
Except for the case $\alpha_1(0)=\alpha_2(0)$ (Figure \ref{fig:dispersive2-h}), where the dispersionless dynamics constructed here is, by Remark \ref{rem:exact-identical}, exact for the compactly-supported initial condition considered, the comparison below should be read in the light of the discussion following Remark \ref{rem:validity}: the dispersionless dynamics is expected to approximate the true dispersionless behaviour, and hence to be relevant for the dispersive dynamics, only in a neighbourhood of the parabolic maxima at $x=0$.
As for the numerical integration of (\ref{eq-Man-NLS}), the initial condition (\ref{eq:disp-initial-condition}) features, in each field, two gradient catastrophes at $x=\pm1$ and hence, as such, it is not suitable for direct numerical implementation. As in the scalar case (see \cite{DORS23}), we found it convenient to use the following infinite product representation of the square-root term in (\ref{eq:disp-initial-condition}):
\begin{equation}
\prod_{j=1}^{\infty}e^{-\frac{x^{2j}}{2j}}=\begin{cases}\sqrt{1-x^2}\, , & -1\leqslant x\leqslant1\, , \\ 0\, , & \text{else}\, .\end{cases}
\end{equation}
Using the above formula, we introduce the following smooth approximations of the initial condition (\ref{eq:disp-initial-condition}):
\begin{equation}
\psi_{k,N}^{(0)}(x)=\exp\left(\frac{i\alpha_k(0)}{2\epsilon}x^2\right)\exp\left(-\sum_{j=1}^{N}\frac{x^{2j}}{2j}\right)\, , \qquad k=1,2\, .
\label{eq:disp-smooth-approx}
\end{equation}
We integrate the focusing vector NLS system (\ref{eq-Man-NLS}) with initial condition (\ref{eq:disp-smooth-approx}) employing a step-adaptive pseudospectral method, following the same procedure and using the same parameters as in \cite{DORS23}. The space derivatives are evaluated using fast Fourier transform routines (FFTW) \cite{FFTW} on a numerical box having size $L=64$, discretised with $2^{17}$ points for each channel, and the time evolution is carried out using the numerical solver included in the ROCK4 C++ library \cite{Abd02}. The code relies on a fourth-order embedded Runge-Kutta method suitable for large stiff problems, adapting each propagation time step to minimise the numerical error and control the scheme stability. We ensured that, at each time, the power spectrum for the spatial Fourier transforms does not reach the boundaries of the spectral domain. As in the scalar case, setting $N=10$ in (\ref{eq:disp-smooth-approx}) is sufficient to ensure a reliable comparison with the semianalytical solution of the dispersionless counterpart. The dispersionless counterpart itself is obtained by directly and numerically integrating (\ref{eq:sigma-system}) via a simple fourth-order embedded Runge-Kutta method with relative accuracy of $10^{-9}$.

Throughout this section we display experiments for the representative value $\epsilon=\frac{1}{10}$, which suffices to illustrate the qualitative comparison with dispersionless predictions. Nonetheless, we expect the dispersive problem to converge to the dispersionless one as $\epsilon$ becomes smaller, as verified in the scalar case.

As illustrated in Figures \ref{fig:dispersive1}--\ref{fig:dispersive-zooms}, the numerical solution of the vector NLS system appears to be driven by the behaviour of the corresponding dispersionless counterpart, exactly as in the scalar problem. Whenever the dispersionless solution is expected to undergo a gradient catastrophe at a finite time $t_c$, the dispersive solution features, in both fields simultaneously -- consistently with Proposition \ref{prop:no-crossing} -- a local maximum for $|\psi_i|$ at a time $t_{\max}$, which is remarkably well approximated by $t_c$, $t_{\max}\approx t_c$. For the catastrophising cases, the dispersionless prediction for the height of $|\psi_i|$ at $x=0$ describes the early time evolution well. When $t$ approaches $t_c$, dispersive effects not captured by the dispersionless model become increasingly relevant, and the two descriptions correspondingly depart from one another. Due to their similarity to the Peregrine soliton and their height, largely exceeding the initial height of the parabolic profile, we interpret these peaks as rogue-wave solutions, exactly as in the scalar case. The presence of additional lobes on the side of the main peak, as in the scalar case, suggests the presence of a high-order rogue wave. The process leading to the formation of these structures qualitatively resembles the Bertola-Tovbis regularisation mechanism of the gradient catastrophe through the generation of a Peregrine breather \cite{BT10,BT13}, exactly as in the semiclassical regularisation of the scalar problem. Lastly, whenever the dispersionless solution is expected to relax, the dispersive solution behaves accordingly, for all the values of the dispersion parameter for which it was possible to carry out the numerical integration.
Based on our broader numerical experience with this problem, beyond the specific cases displayed here, we observe that the competition between the two known semiclassical regularisation mechanisms of the gradient catastrophe -- the generation of a single Peregrine-like breather (Bertola-Tovbis scenario) and the generation of a lattice of counterpropagating breathers coalescing into a single, higher central peak, as observed in the dam-break problem (El-Khamis-Tovbis scenario) \cite{ElKT16} -- also takes place in the vector case, exactly as conjectured for the scalar problem. Specifically, when the initial chirps $\alpha_1(0),\alpha_2(0)$ lie in the gradient catastrophe region of the $(\alpha_1(0),\alpha_2(0))$-plane, sufficiently close to the infinite, continuous branch of the threshold curve $\Delta_Q=0$, the observed dynamics is reminiscent of the El-Khamis-Tovbis scenario, with the dispersionless prediction of the catastrophe time correspondingly less accurate -- naturally so, since the interacting breathers need to coalesce before the central peak is reached. Moving away from that branch, while remaining in the catastrophe region, the mechanism of formation of the central peak instead resembles the Bertola-Tovbis scenario, as is the case for all the examples displayed in this section. This qualitative picture is, at present, a conjecture based on numerical evidence alone, exactly as in the scalar case, where an analytical explanation is likewise not yet available.
The numerical evidence, supporting the analytical predictions from the dispersionless model, suggests that the coupled rogue wave formation (or its suppression) for the focusing vector NLS system with initial condition (\ref{eq:disp-smooth-approx}) can be controlled and manipulated by conveniently tuning the two chirps and the shared amplitude of the initial parabolic profiles, exactly as observed for the scalar problem. \colorblue{More significantly, the dispersionless model remains a good predictor of the fully dispersive dynamics even beyond the vanishing-dispersion regime: this is already the case at $\epsilon=1/10$ for all the examples displayed here and, in our broader numerical experience, for larger values as well. A systematic characterisation of how this agreement degrades as $\epsilon$ increases lies beyond the scope of the present work.}

\begin{figure}[H]
\centering
\begin{subfigure}{\textwidth}
\centering
\includegraphics[width=\textwidth]{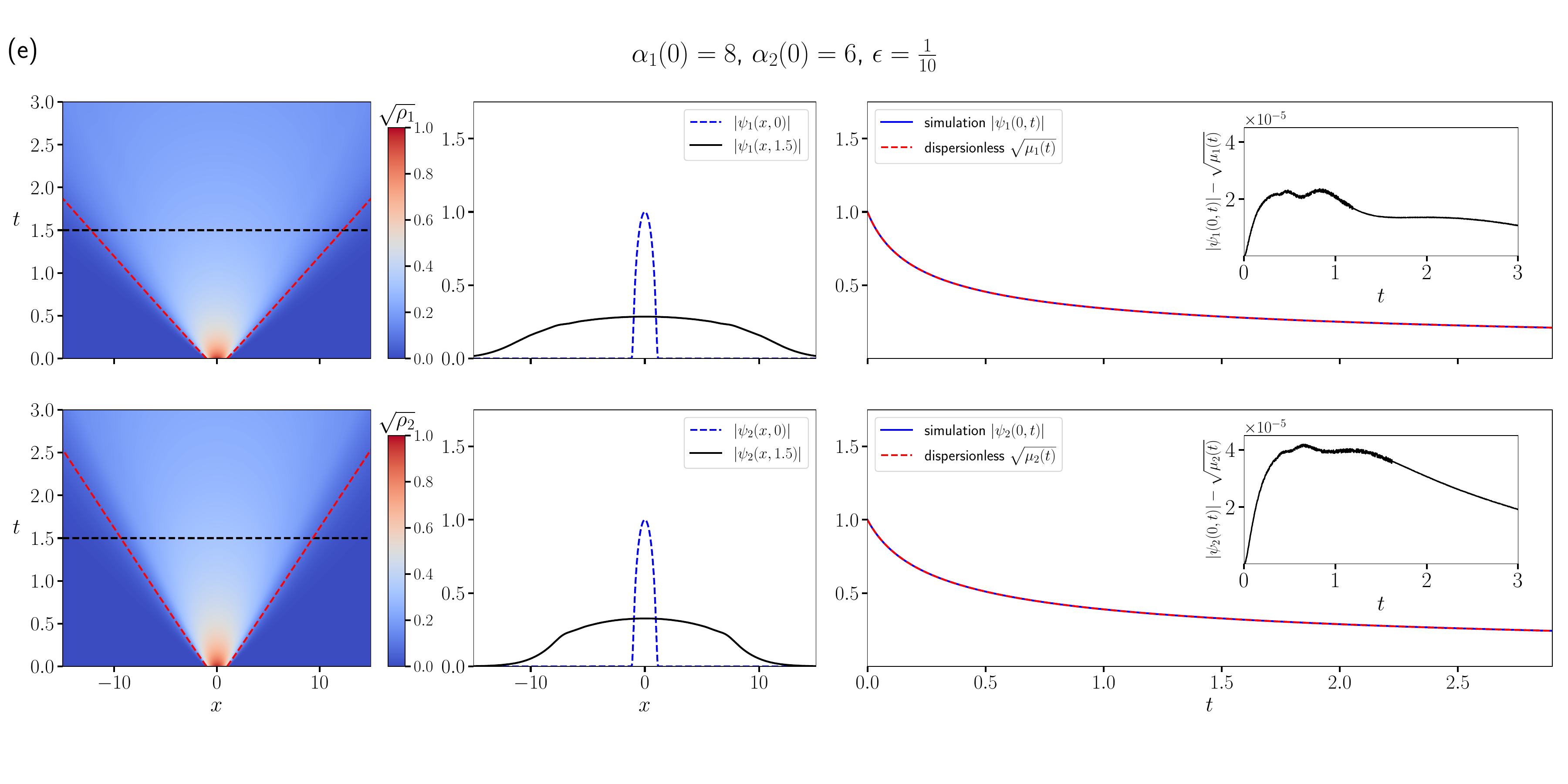}\setcounter{subfigure}{4} 
\captionsetup{labelformat=empty}
\caption{}
\label{fig:dispersive1-e}
\end{subfigure}
\begin{subfigure}{\textwidth}
\centering
\vspace{-1cm}\includegraphics[width=\textwidth]{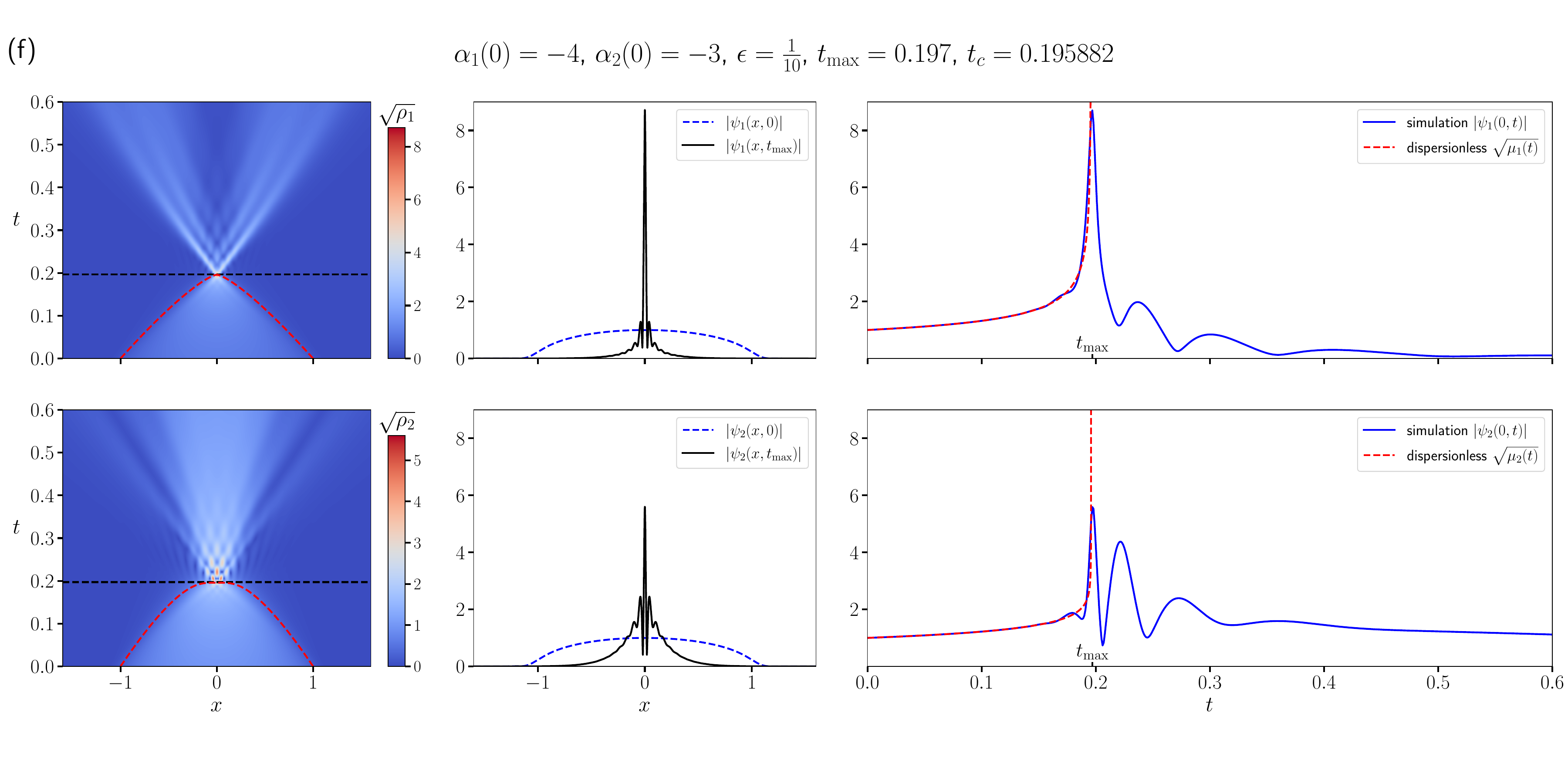}
\captionsetup{labelformat=empty}
\caption{}
\label{fig:dispersive1-f}
\end{subfigure}
\vspace{-1.5cm}\caption{Numerical evolution of the initial condition $\psi_{1}(x,0)$, $\psi_{2}(x,0)$ given by (\ref{eq:disp-initial-condition}) for $\epsilon=\frac{1}{10}$, for two of the four cases marked in Figure \ref{fig:region-plot-symmetric}. In each group of panels, the first row illustrates the evolution of the first field, whereas the second row illustrates the evolution of the second field. In each row, the panel on the left shows the evolution of the absolute value of $\psi$ as an $(x,t)$-plot, where the dashed black line indicates the time of the maximum $t_{\max}$ and the dashed red line indicates where the dispersionless densities vanish, $x=-L_{i}(t)$ and $x=L_{i}(t)$, as in (\ref{eq:support}), treated here as the end points of the dispersionless support; the panel in the centre shows the initial condition as a dashed blue line and the absolute value of $\psi$ at $t = t_{\max}$ as a solid black line; the panel on the right shows the numerical evolution of the absolute value of $\psi$ at $x = 0$ for all values of $t$ in the range of numerical integration as a solid blue line, overlapping the analytical evolution of the corresponding dispersionless solution at $x = 0$ [namely, $\sqrt{\mu_i(t)}$] as a dashed red line. It is possible to observe that, as the dispersionless solution features a blowup at finite time $t_c$, the corresponding dispersive numerical solution features a local maximum at time $t_{\max}$, where $t_c$ and $t_{\max}$ are in excellent agreement. \textit{Figure \ref{fig:dispersive1-e}}: $\alpha_1(0)=8,\alpha_2(0)=6$, corresponding to point e in Figure \ref{fig:region-plot-symmetric}, where both fields relax, consistently with the dispersionless prediction; the height of $|\psi_i|$ at $x=0$ is in excellent agreement with the dispersionless prediction over a long time interval. \colorblue{Note that this case lies in the same domain of asymptotic relaxation as point b in Figure \ref{fig:region-plot-symmetric}, so Figure \ref{fig:dispersive1-e} illustrates the dispersive counterpart of the regime shown in Figure \ref{fig:regime-b}.} \textit{Figure \ref{fig:dispersive1-f}}: $\alpha_1(0)=-4,\alpha_2(0)=-3$, corresponding to point f in Figure \ref{fig:region-plot-symmetric}, where both fields develop a rogue-wave-like peak at $t_{\max}\approx t_c$, in excellent agreement with the dispersionless catastrophe time; the smaller of the two chirps drives the coupled system to an earlier $t_c$ than either the second channel alone (decoupled) or the case of two equal chirps would give.}
\label{fig:dispersive1}
\end{figure}

\begin{figure}[H]
\ContinuedFloat
\centering
\begin{subfigure}{\textwidth}
\centering
\includegraphics[width=\textwidth]{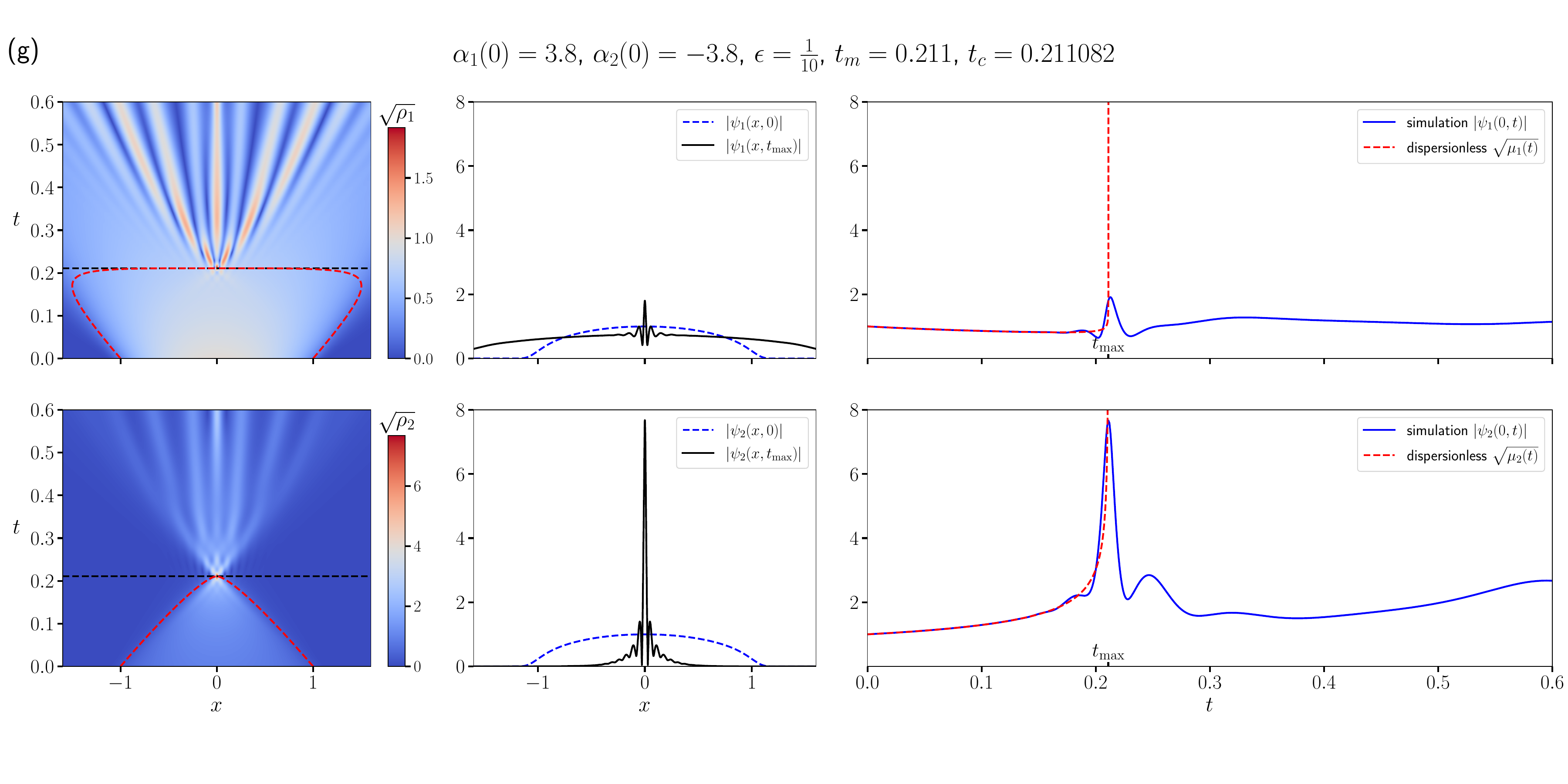}
\captionsetup{labelformat=empty}
\caption{}
\label{fig:dispersive2-g}
\end{subfigure}
\begin{subfigure}{\textwidth}
\centering
\vspace{-1cm}\includegraphics[width=\textwidth]{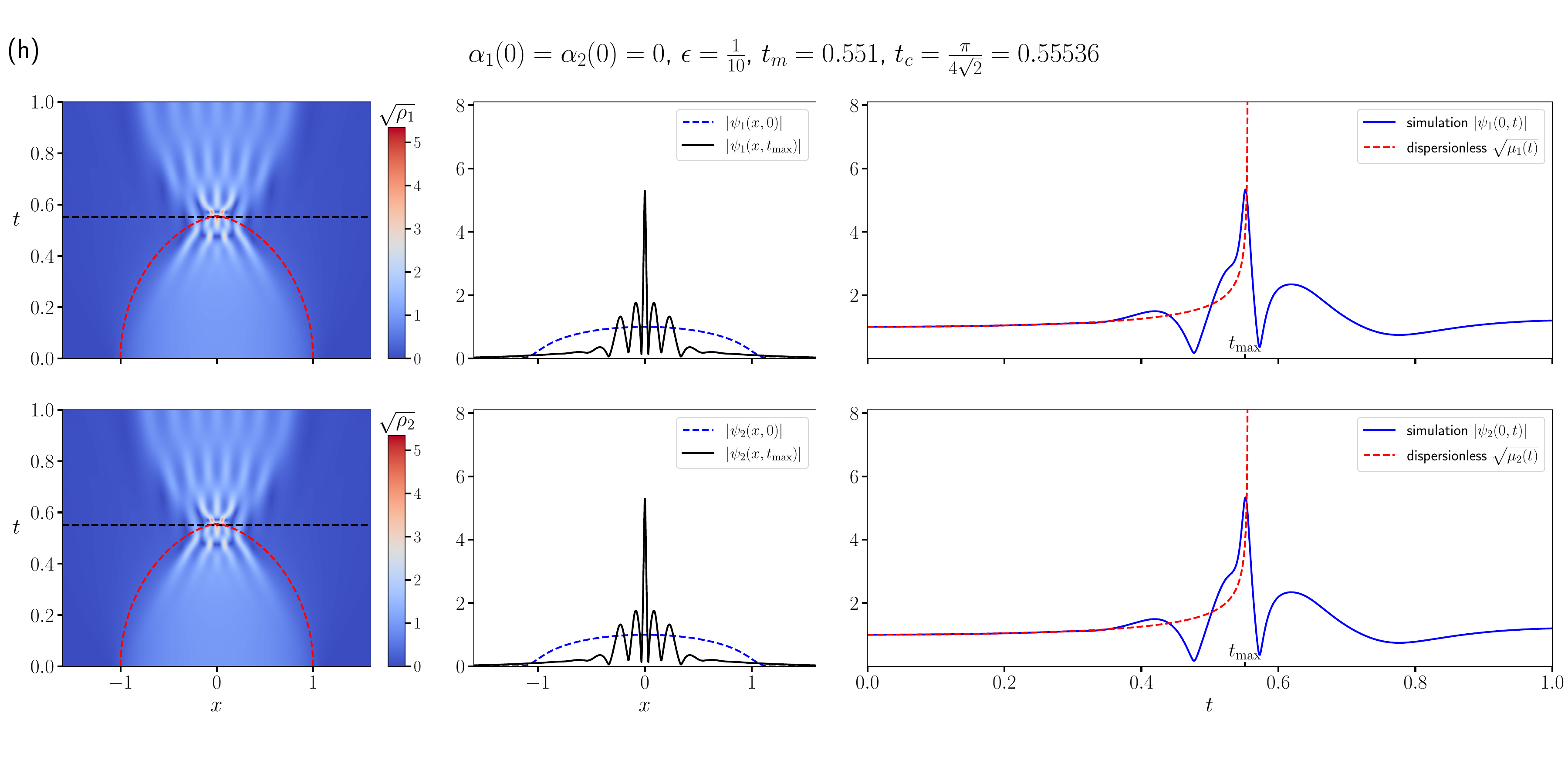}
\captionsetup{labelformat=empty}
\caption{}
\label{fig:dispersive2-h}
\end{subfigure}
\vspace{-1.5cm}\captionsetup{labelsep=space}\caption{(continued). Numerical evolution of the initial condition $\psi_{1}(x,0)$, $\psi_{2}(x,0)$ given by (\ref{eq:disp-initial-condition}) for $\epsilon=\frac{1}{10}$, for two of the four cases marked in Figure \ref{fig:region-plot-symmetric}. In each group of panels, the first row illustrates the evolution of the first field, whereas the second row illustrates the evolution of the second field. In each row, the panel on the left shows the evolution of the absolute value of $\psi$ as an $(x,t)$-plot, where the dashed black line indicates the time of the maximum $t_{\max}$ and the dashed red line indicates where the dispersionless densities vanish, $x=-L_{i}(t)$ and $x=L_{i}(t)$, as in (\ref{eq:support}), treated here as the end points of the dispersionless support; the panel in the centre shows the initial condition as a dashed blue line and the absolute value of $\psi$ at $t = t_{\max}$ as a solid black line; the panel on the right shows the numerical evolution of the absolute value of $\psi$ at $x = 0$ for all values of $t$ in the range of numerical integration as a solid blue line, overlapping the analytical evolution of the corresponding dispersionless solution at $x = 0$ [namely, $\sqrt{\mu_i(t)}$] as a dashed red line. It is possible to observe that, as the dispersionless solution features a blowup at finite time $t_c$, the corresponding dispersive numerical solution features a local maximum at time $t_{\max}$, where $t_c$ and $t_{\max}$ are in excellent agreement. \textit{Figure \ref{fig:dispersive2-g}}: $\alpha_1(0)=3.8,\alpha_2(0)=-3.8$, corresponding to point g in Figure \ref{fig:region-plot-symmetric}, where the coupling drives both fields into a simultaneous rogue-wave-like peak at $t_{\max}\approx t_c$, even though, decoupled, the first field would have relaxed on its own. \textit{Figure \ref{fig:dispersive2-h}}: $\alpha_1(0)=\alpha_2(0)=0$, corresponding to point h in Figure \ref{fig:region-plot-symmetric}, showing the same simultaneous peak formation at $t_{\max}\approx t_{c}=\pi/(4\,\sqrt{2})$ for identical chirps; here the two identical fields reinforce each other through the coupling, so that $t_c$ occurs earlier than for a single, decoupled channel with the same chirp. As in the scalar case, the dashed red line marking the dispersionless support is comparatively less accurate for this particular case. Since $\alpha_1(0)=\alpha_2(0)$ and the two initial supports coincide, the dispersionless dynamics shown here is, by Remark \ref{rem:exact-identical}, the exact dispersionless dynamics for this compactly-supported initial condition.}
\label{fig:dispersive2}
\end{figure}

\begin{figure}
\centering
\begin{subfigure}{0.32\textwidth}
\centering
\includegraphics[width=\textwidth]{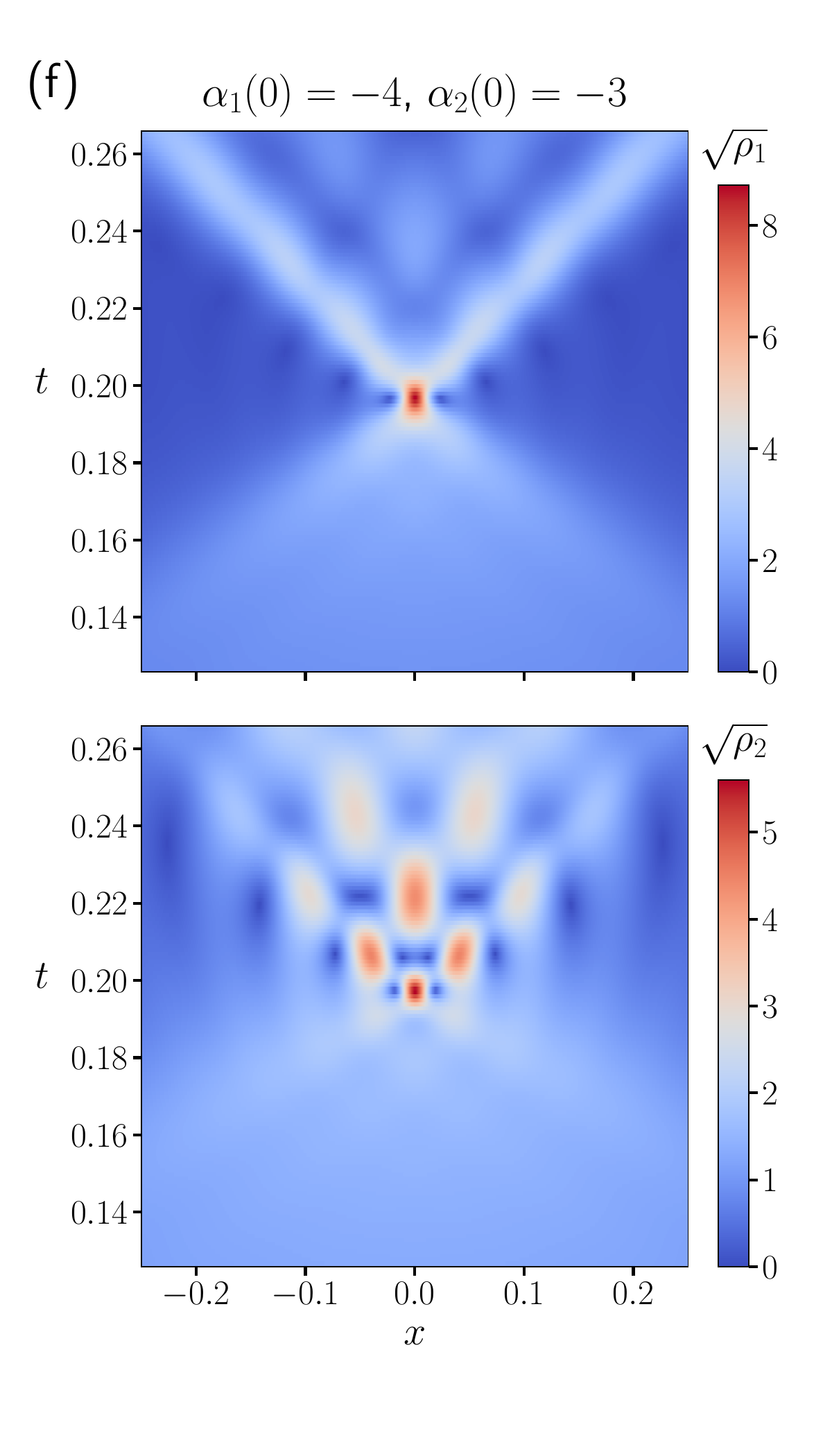}\setcounter{subfigure}{5} 
\captionsetup{labelformat=empty}
\caption{}
\label{fig:dispersive-zooms-f}
\end{subfigure}
\hfill
\begin{subfigure}{0.32\textwidth}
\centering
\includegraphics[width=\textwidth]{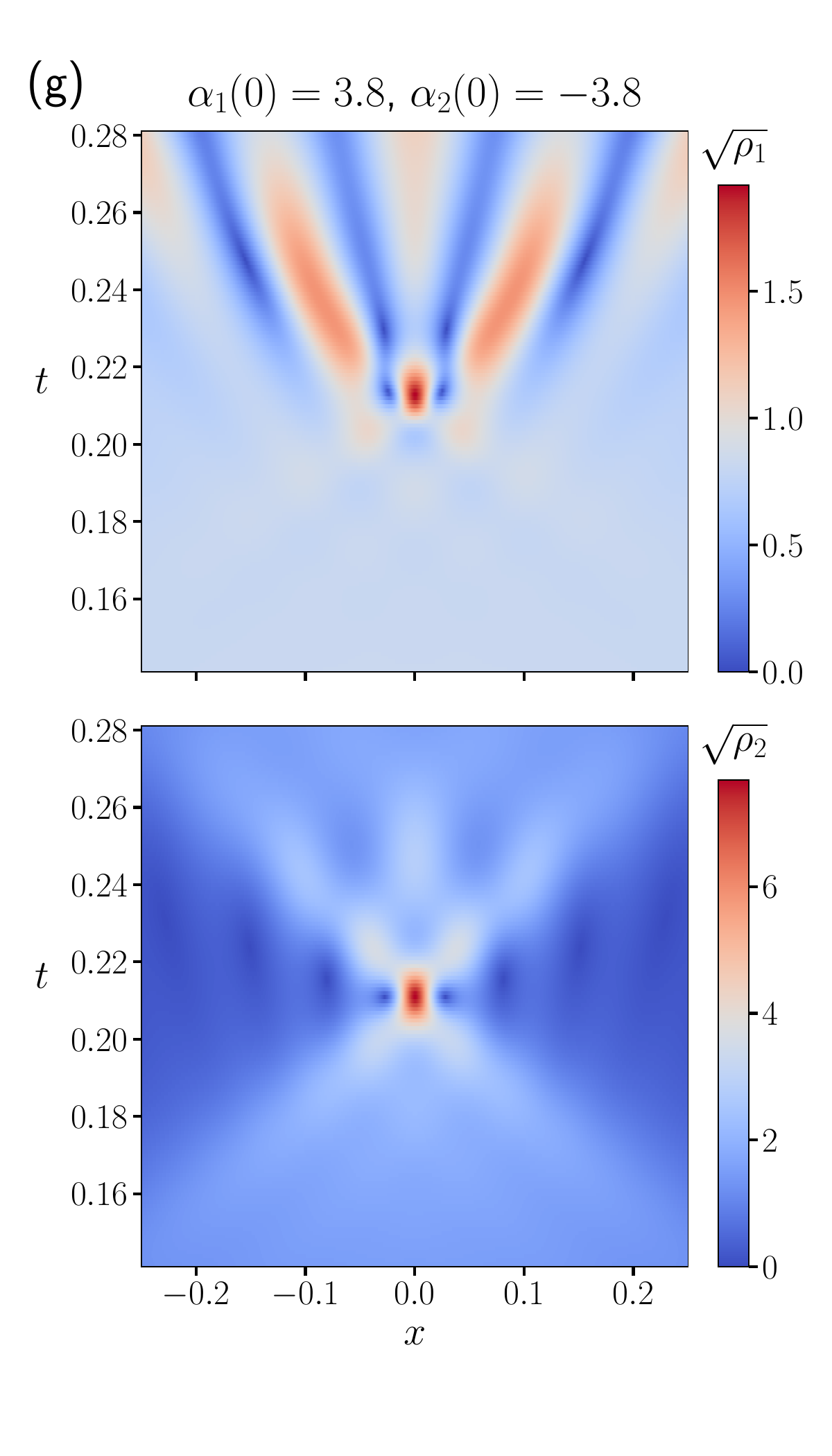}
\captionsetup{labelformat=empty}
\caption{}
\label{fig:dispersive-zooms-g}
\end{subfigure}
\hfill
\begin{subfigure}{0.32\textwidth}
\centering
\includegraphics[width=\textwidth]{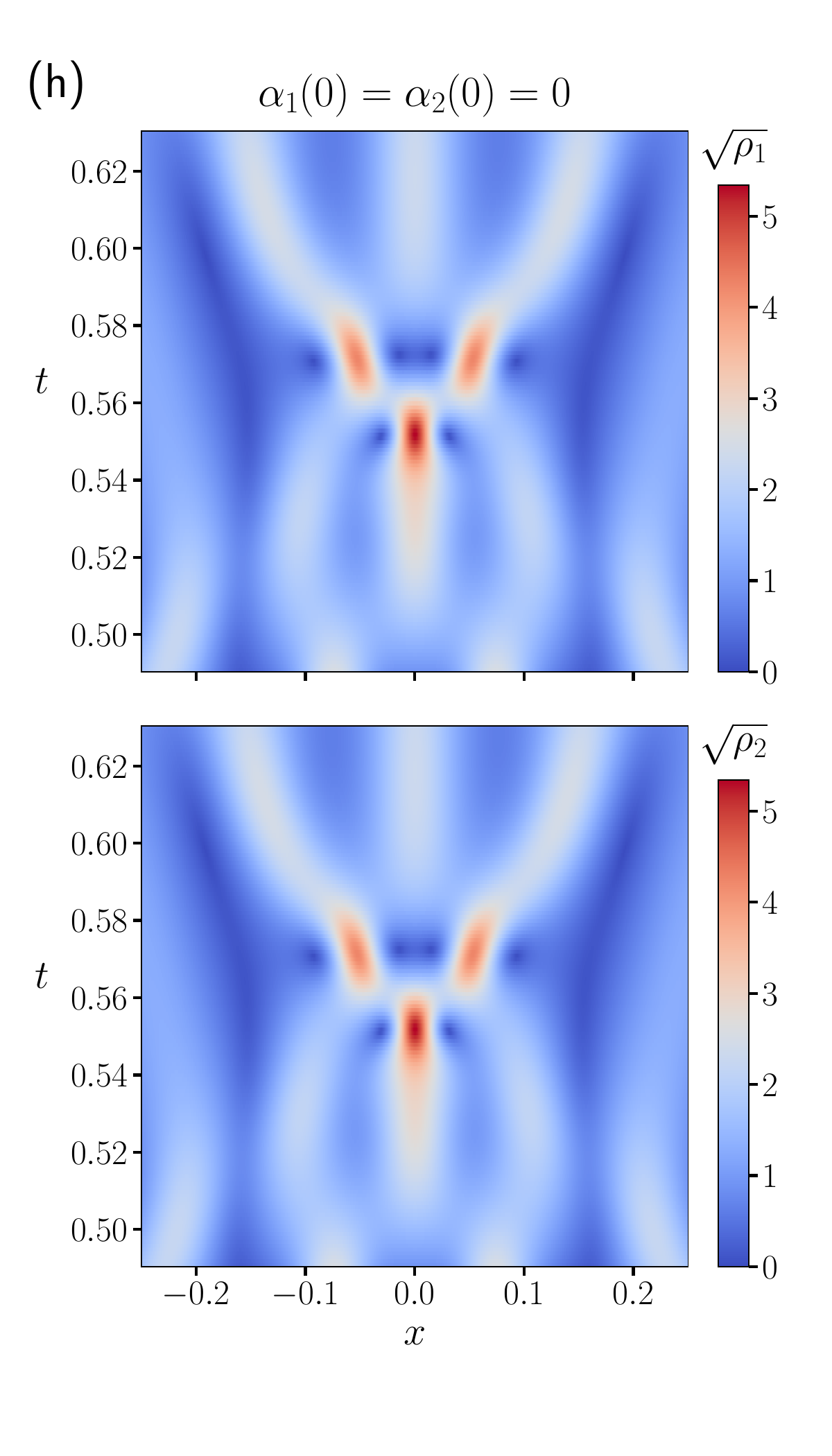}
\captionsetup{labelformat=empty}
\caption{}
\label{fig:dispersive-zooms-h}
\end{subfigure}
\caption{Magnification of the $(x,t)$-plot around the peaks for each one of the cases in Figure \ref{fig:dispersive2}, showing a rogue-wave-like behaviour, with two troughs neighbouring the maximum, as in the Peregrine soliton. Figures \ref{fig:dispersive-zooms-f}, \ref{fig:dispersive-zooms-g} and \ref{fig:dispersive-zooms-h} refer to the corresponding Figures \ref{fig:dispersive1-f}, \ref{fig:dispersive2-g} and \ref{fig:dispersive2-h}, respectively.}
\label{fig:dispersive-zooms}
\end{figure}
\section{Conclusions and Outlooks}
\label{sec-conclusions}

In this paper we have studied the focusing vector nonlinear Schr\"odinger equation, or Manakov system, for a pair of coupled initial conditions with compactly supported, concave parabolic density profiles and phases depending quadratically on the spatial coordinate. Recasting the dispersionless limit of the system in fluid-dynamical form, we have shown that it is elliptic wherever both densities are strictly positive, degenerating to a parabolic type at the vacuum points of either field, and that it admits a Hamiltonian formulation with respect to a natural set of conserved densities. 

Focusing on the natural class of self-similar solutions generated by such compactly-supported, parabolic-profile initial data – what we have termed Talanov dynamics – we extend the results obtained for the scalar single field to the Manakov two-field case. For arbitrary initial chirps in the two fields, we provide closed-form, implicit expressions for the evolution of the two fields, and an explicit expression for the time of the gradient catastrophe, whenever it occurs.

For identical initial supports, we have proved that the two supports never cross throughout the evolution, and classified the effect of the coupling on the catastrophe\colorblue{. Depending} on the initial chirps, the two fields may catastrophise, or relax, in agreement with what each field would do if decoupled\colorblue{. However, coupling} can also force a simultaneous catastrophe in a field that would otherwise relax on its own, or even in both fields when both would relax if decoupled. Whenever catastrophe occurs in the coupled system, we have observed numerically that it occurs earlier than it would in either decoupled field. For identical initial supports, and in the presence of dispersion, we have numerically shown that the same dispersionless criterion determines, in the semiclassical regime, whether the fully dispersive solution relaxes or develops a rogue wave in each field, whose onset time is accurately predicted by the corresponding dispersionless catastrophe time, in direct analogy with the scalar case.

The dispersionless dynamics studied here for compactly-supported, parabolic initial data is exact only when both the initial chirps and the initial supports coincide (Remark \ref{rem:exact-identical}). Otherwise, it is expected to be an approximation only in a neighbourhood of the parabolic maxima, under the assumption that the two supports remain overlapping in that neighbourhood. Nonetheless, to the best of our knowledge, we have provided here, for the first time, a criterion for the formation of a peak (rogue wave) in the time evolution, or for asymptotic relaxation, that is valid for the fully dispersive dynamics, and that we have verified a posteriori against direct numerical integration, in the semiclassical, small-dispersion regime. A more rigorous approach, beyond the numerical evidence gathered here, is the subject of future research.

A rigorous, analytical account of the mechanisms responsible for the emergence of a rogue wave in the dispersive system, of the kind already available for a related class of scalar initial data \cite{BJM25}, would place the numerical evidence gathered here, and in the scalar case \cite{DORS23}, on a firmer footing. More broadly, the Talanov self-similar framework developed here for the vector NLS system invites extension to more general settings -- involving, for instance, different or more general forms of coupling between the fields, or media whose properties vary in space and/or time, an area of growing interest in modern photonics \cite{Gal22}. We leave a systematic exploration of these directions to future work.

\subsubsection*{Acknowledgments}
The authors are grateful to Fabio Baronio, Costantino De Angelis, Gennady El, Evgeny Ferapontov, Theodoros Horikis, Karima Khusnutdinova, Sara Lombardo, Antonio Moro, Maria Grazia Naso, Elena Vuk, and Federico Zullo  for useful and clarifying discussions.
We also gratefully acknowledge the auspices of the GNFM Section of INdAM, under which part of this work was carried out, and the financial support of the project MMNLP (Mathematical Methods in Non Linear Physics) of the INFN.
\subsubsection*{Conflict of Interest Statement}
The authors declare no conflict of interest.
\bibliographystyle{unsrt}
\bibliography{Manakov_References}
\end{document}